Technische Universität Braunschweig
Carl-Friedrich-Gauß-Fakultät, Department Informatik
Institut für Software Systems Engineering

# System Model Semantics of Statecharts

Informatik-Bericht 2008-04

**María Victoria Cengarle**[1], **Hans Grönniger**[2]
and **Bernhard Rumpe**[2]
with the help of
**Martin Schindler**[2]

[1]Software and Systems Engineering,
Technische Universität München, Germany
[2]Software Systems Engineering,
Technische Universität Braunschweig, Germany

Software
Systems
Engineering

July 8, 2008

# Contents

In this report, semantics for Statecharts is defined based on a mathematical model of object systems called “system model”. The semantics definition is detailed for UML/P Statecharts, a variant of Statecharts which restricts the use of a few methodologically and semantically difficult concepts. After transforming full UML/P Statecharts to simplified but semantically equivalent Statecharts, the semantics is defined denotationally as a mapping into the system model. It is also sketched how already existing Statechart semantics can be mapped into the system model. This report follows [4], in which we introduced our approach in detail and defined semantics for UML class diagrams.

# 1 Introduction

In this report, we give a semantics for UML/P Statecharts [11], a variant of UML StateMachines [10] which restricts the use of a few methodologically and semantically difficult concepts.

We follow our approach for the semantics definition as outlined in [4].

Semantics of UML/P Statecharts is defined in two steps. Full UML/P Statecharts are transformed into simplified but semantically equivalent UML/P Statecharts in the fist step. By that transformation, the number of syntactical concepts used can be reduced significantly without losing expressiveness. In the second step, simplified UML/P Statecharts are given semantics by mapping them into the system model which is a general mathematical model of object systems that constitutes the semantic domain for Statecharts as well as other UML diagram types. We assume that the reader is familiar with the system model definitions which can be found in [1, 2, 3].

First, in Chapter 2, the concrete syntax as MontiCore grammars is specified. In Chapter 3, we introduce the equivalent mathematical abstract syntax of full UML/P Statecharts. In Chapter 4, we state context conditions that must hold for well-formed Statecharts. After that, we provide transformation rules to obtain conceptually simplified Statecharts from full UML/P Statecharts in Chapter 5. After applying the transformations, we may syntactically simplify the Statecharts that leads to a simplified abstract syntax of UML/P Statecharts given in Chapter 6. The latter allows for a more concise and understandable semantics mapping in Chapter 7. In Chapter 8, we show how already existing Statechart semantics can be mapped to the system model in general and give a concrete example by mapping the semantics of [13]. The last chapter discusses related work and concludes the paper.

# 2 Concrete Syntax of Full UML/P Statecharts

The concrete syntax of UML/P Statecharts can be found below defined in the MontiCore grammar format [1]. Concrete examples, and a comparison to UML State Machines can be found in [11] which introduces this methodologically and semantically improved variant of Statecharts in detail.

The first grammar `SC.mc` is complete, only comments and some technical details regarding parser generation have been removed. The second grammar is an excerpt of `Common.mc` where only the relevant definitions are shown. `Common.mc` is a grammar that is the super grammar of all UML/P grammars providing frequently used concepts such as stereotypes.

MontiCore-Grammar

```
package mc.umlp.sc;

/**
@version 1.0
*/
grammar SC extends mc.umlp.common.Common {

  external Statements;
  external Expression;

  interface SCElement;
  interface SCEvent;

  SCDefinition =
    Completeness?
    Stereotype?
    "statechart" Name:IDENT
    (SCMethod | ClassName:ClassOrInterfaceType)?
    "{" (Invariants:Invariant ";" | SCElements:SCElement)* "}";

  SCMethod =
    Name:QualifiedName
      "(" (SCParameters:SCParameter ("," SCParameters:SCParameter)*)? ")";

  SCParameter = Type Name:IDENT;

  SCAction =
    (PreCondition:Invariant)?
      (";" | ("/" Statements (PostCondition:Invariant ";")?);

```

[1]The MontiCore UML/P grammars, version 1.0, were developed by Martin Schindler as part of the MontiCore project (www.monticore.org).

```
  SCDoAction = "do" SCAction;

  SCEntryAction = "entry" SCAction;

  SCExitAction = "exit" SCAction;

  SCModifier =
    Stereotype?
    (Initial:["initial"] | Final:["final"] | Local:["local"])*;

  SCState implements SCElement =
    Completeness?
    SCModifier
    "state" Name:IDENT
    (("{"
       (Invariant ";")?
        SCEntryAction? SCDoAction? SCExitAction?
        (SCInternTransitions:SCInternTransition |
         SCElements:SCElement)*
     "}") | ";");

  SCInternTransition = Stereotype? "->" ":"? SCTransitionBody;

  SCTransition implements SCElement =
    Stereotype?
    Source:QualifiedName "->" Target:QualifiedName
    ((":" SCTransitionBody) | ";");

  SCTransitionBody =
    (PreCondition:Invariant)?
    SCEvent?
        (("/" Statements (PostCondition:Invariant ";")? | ";");

  SCMethodOrExceptionCall implements SCEvent =
    Name:QualifiedName (SCArguments)?;

  SCReturnStatement implements SCEvent =
    "return"
      ("("
        Incomplete:[INCOMPLETE:"..."]
        | Expression
      ")")?;

  SCArguments =
    ("(" Incomplete:[INCOMPLETE:"..."] ")")
    | ("(" ")")
    | ("(" Expressions:Expression ("," Expressions:Expression)* ")");

  SCCode implements SCElement = "code" Statements;
}
```

MontiCore-Grammar

```
package mc.umlp.common;

/**
@version 1.0
*/
grammar Common {

  ident IDENT =
    ('a'..'z' | 'A'..'Z' | '_' | '$')
    ('a'..'z' | 'A'..'Z' | '_' | '0'..'9' | '$')*;

  Stereotype =
    "<<" Values:StereoValue ("," Values:StereoValue)* ">"">";

  StereoValue = Name:IDENT ("=" Value:STRING)?;

  QualifiedName = Names:IDENT "." Names:IDENT)*;

  ClassOrInterfaceType = Name:QualifiedName TypeArguments?;

  TypeArguments =
    "<" TypeArguments:TypeArgument ("," TypeArguments:TypeArgument)* ">";

  TypeArgument =
    Type
    | ("?" (("extends" UpperBound:ReferenceType) |
            ("super" LowerBound:ReferenceType))?);

  Invariant =
    (Kind:IDENT ":")?
    "[" InvariantExpression(parameter Kind) "]";

  external InvariantExpression;

  interface Type;
  interface ReturnType;

  PrimitiveType implements ReturnType, Type =
    Primitive: [ "boolean" | "byte" | "char" | "short"
               | "int" | "float" | "long" | "double"] ("[""]")*;

  ReferenceType implements ReturnType, Type =
    ClassOrInterfaceType ("[""]")*;

  Completeness = ...

  // ...
}
```

# 3 Abstract Syntax of Full UML/P Statecharts

We proceed with defining the mathematical abstract syntax of UML/P Statecharts. Please note, that we have refactored the syntax. For example, we renamed rules for a more comprehensible reference later on, and we also changed details of the language as follows:

- As explained in [11, Sect. 3.4], completeness information is a syntactical means to indicate that model or model element is a view (other models might exist to complete the model) or regarded as the complete model of the system. This does not constrain the actual system from have additional behavior or structure. Hence, completeness information is irrelevant for the semantics and removed from the syntax, e.g., lines 12 and 39 in grammar `SC.mc`.
- In principle, arbitrary stereotypes can be defined for Statecharts, states, or transitions. For the semantics, we only consider the stereotypes explained in [11] and fix them in the abstract syntax. Additional stereotypes (also for other model elements) can of course later be added and their meanings incorporated in the semantics.
- In line 15 of the grammar `SC.mc`, the detailed type specification is replaced by a simple name, assuming that qualified class names can be somehow encoded. Support for generic types is not considered in this version of the semantics but is a matter of future investigations.
- Additionally, the semantics does not consider method Statecharts (again, line 15), rules regarding method Statecharts have also been removed.
- In line 16 of `SC.mc`, multiple invariants are allowed. We assume them connected by "and" and only use one invariant.
- Arbitrary additional code in Statecharts (line 76) is not allowed in the mathematical variant. This is also a matter of further investigations.
- The interface SCElements in line 8 has been expanded as sets of transitions and states in the mathematical syntax.
- We removed the modifier local (line 36) as its intended use refers only to syntax: the local element may not be referenced in other diagrams.
- States do not directly contain substates or transitions in the mathematical version. We use a relation "sub" instead, that is the substate relation. That means, SCElements can be removed in line 45. As a consequence, all states and transitions are kept in SCFull. This implies that state names have to be unique, and no qualified names in transition source or target are needed.

- We renamed SCEvent to Call, and also assume that calls are given as a set from an (action) language, i.e., SCMethodOrExeceptionCall, SCReturnStatement, and SCArguments can be removed (lines 61 - 74).
- Invariants in `Common.mc` (line 26) are parameterized with the name of the invariant language. This is solution to a technical problem and not needed in the mathematical version.
- Identifiers are specified in `Common.mc`, we assume a set of Names which is left unspecified.

Some minor changes and rearrangements such as renaming (SCDefinition to SCFull, IDENT to Name) will not be further detailed.

The abstract syntax for Statecharts is given below in mathematical form. We use the set *Name* that contain identifiers which are not further specified. The sets *Call*, *Cond*, and *Stmt* are also not described here but specified elsewhere. We import those parts of the language in a compositional form as described in [8]. They are part of an action language that may for example be OCL to state conditions or Java to formulate statements, calls, or other expressions like conditions. In order to obtain an unambiguous variable assignment, the arguments of a call expression should only be "constructor" expressions, like $i + 1$ or $a : as$ where $a$ is head and $as$ rest of a list in case of a typical functional language.

$$
\begin{array}{lcl}
\mathit{SCFull} & = & \wp(\mathit{SCStereo}) \times \mathit{DiagramName} \times \mathit{ClassName} \times \mathit{Inv}^{opt} \times \\
 & & \wp(\mathit{State}) \times \wp(\mathit{Trans}) \times \mathit{Sub} \\
\mathit{SCStereo} & = & \{\textsf{prio:inner}, \textsf{prio:outer}, \textsf{completion:ignore}, \\
 & & \ \textsf{completion:chaos}, \textsf{action conditions:sequential}\} \\
\mathit{Sub} & = & \wp(\mathit{StateName} \times \mathit{StateName}) \\
\mathit{State} & = & \wp(\mathit{SStereo}) \times \wp(\mathit{Modifier}) \times \mathit{StateName} \times \mathit{Inv}^{opt} \\
 & & \mathit{Entry}^{opt} \times \mathit{Exit}^{opt} \times \mathit{Do}^{opt} \times \wp(\mathit{InternT}) \\
\mathit{SStereo} & = & \{\textsf{error}, \textsf{exception}\} \\
\mathit{Modifier} & = & \{\textsf{initial}, \textsf{final}\} \\
\mathit{InternT} & = & \mathit{Pre}^{opt} \times \mathit{Call} \times \mathit{Act}^{opt} \\
\mathit{Trans} & = & \mathit{TStereo}^{opt} \times \mathit{Src} \times \mathit{Pre}^{opt} \times \mathit{Call} \times \mathit{Act}^{opt} \times \mathit{Trg} \\
\mathit{TStereo} & = & \{\textsf{prio = n}\}, \text{ where } n \in \mathbb{N} \\
\mathit{Inv}, \mathit{Pre} & = & \mathit{Cond} \\
\mathit{Act}, \mathit{Entry}, \mathit{Exit}, \mathit{Do} & = & \mathit{Stmt} \times \mathit{Cond}^{opt} \\
\mathit{Src}, \mathit{Trg} & = & \mathit{StateName} \\
\mathit{DiagramName}, & & \\
\mathit{ClassName}, & & \\
\mathit{StateName} & = & \mathit{Name}
\end{array}
$$

For notational convenience, we refer to specific components of these tuples using their (dot separated) names. If the component is a set, we use the plural form of the name. E. g., given a Statechart $sc \in SCFull$, $sc.diagramName$ is a shorthand for the projection on the second component: $sc.diagramName = \pi_2(sc)$, and $sc.states$ denotes a projection on the fifth: $sc.states = \pi_5(sc)$.

# 4 Context Conditions for Full UML/P Statecharts

Well-formed Statecharts must fulfill context conditions that are listed below. Transformation rules and semantics are only defined for well-formed Statecharts.

Please note that this is not a complete list. Context conditions are handled together with the language definition using the MontiCore framework [9] and are not treated extensively in this report.

1. The transitive closure of *Sub* is irreflexive. No state is substate of itself.
2. If a transition trigger is an exception, a state with stereotype exception must exist.
3. At most one priority stereotype and completion stereotype may be used. If a completion stereotype exists, the Statechart may not contain an error state.
4. Source and target states of a transition exist, i. e., are declared in the Statechart.
5. The class name is declared in the underlying class diagram.
6. A method called or signal raised (cf. *Event*) is likewise declared in the underlying class diagram.
7. The formal parameters of an event are pairwise different.
8. The invariants may refer only to reachable attributes or query methods, starting from the object in question.
9. Pre- and postconditions can additionally refer to the arguments of the event.
10. Statements may not contain method calls that (eventually) lead to events that are triggers for the Statechart.
11. The statements refer only to reachable attributes or methods starting from the object.
12. States have all different state names.
13. Initial states that have an outgoing transition which is a constructor call may not have ingoing transitions.
14. Final states that have an ingoing transition which is a finalize method call may not have outgoing transitions.

# 5 Transformation of Full UML/P Statecharts

In order to obtain a semantics for Full UML/P Statecharts, we fist define transformation rules that transform Full UML/P Statecharts into simplified UML/P Statecharts which are them mapped into the system model. We claim that the transformation rules described hereafter preserve well-formedness and semantics of the Statechart. The whole transformation consists of 26 transformation rules, expressed as relations over full UML/P Statecharts.

## 5.1 Transformation Scheme

A single transformation rule is structured by 6 compartments as shown in the following example scheme 1:

| 1. transformationName | |
|---|---|
| bind | variable bindings |
| pre | preconditions |
| $\Delta$ | elements that change |
| trafo | actual transformation |
| comment | additional explanations |
| example | example by figure or text |

The first compartment (bind) contains variable bindings. Bound variables may be referenced in the following compartments. The second compartment (pre) contains preconditions for the transformation rule. The $\Delta$ compartment specifies which variables can be changed by the transformation. All variables not mentioned here stay the same. The effect of the transformation is described in the fourth compartment (trafo) where altered variables are marked with a prime. In the fifth compartment, a comment may be provided, and in the last one an example may be given. All transformations are given a number and a name for later reference.

Given a transformation rule $R$, two Statecharts $sc$ and $sc'$ are in a transformation relation, $(sc, sc') \in R$ if the transformation rule is applicable (i.e., the precondition holds) and $sc'$ is the same as $sc$ except for the variables stated in $\Delta$ which are altered according to the trafo compartment.

## 5.2 Assumptions and Helpers

In the following, we state which assumptions were made when defining the transformations and also which helper functions, relations, or conditions are used.

**Assumptions:**

| Name | Definition |
|---|---|
| *setTimer*, *stopTimer*, *timeout* | *setTimer*, *stopTimer*, *timeout* $\in$ *STMT* are elements of *STMT* that represent statements that can be used to model a timer. Some framework functionality is assumed that implements timers correctly. |
| & | Statements can be concatenated. $s_1, s_2 \in \mathit{STMT} \implies s_1\&s_2 \in \mathit{STMT}$ |
| !, && *true*, *false* | Common logic operations exist for *Cond*, e.g., negation (!), conjunction (&&) etc. We use $\&\&_{1\leq i\leq n}\ c_i$ as a shortcut for $c_1\ \&\& \ldots \&\&\ c_n$. *true*, *false* $\in$ *Cond* represent the boolean values true and false. |
| $+$ | Actions are made up of statements and (optional) postconditions. For instance, an entry action $e$ is a pair $(\mathit{act}, \mathit{post}) \in \mathit{STMT} \times \mathit{COND}$. As stated above, given two actions $e_1, e_2$, we write $e_1$*.act* & $e_2$*.act* to concatenate the statements, $e_1$*.post* && $e_2$*.post* for the conjunction of postconditions Additionally, we write $e_1 + e_2$ to form a new action: We assume for $+$ that the conditions can be *sequentially interleaved* with the statements and that an appropriate mechanism exists that handles cases in which a condition in the action sequence evaluates to false. |
| *nameOf* | There is an operation *nameOf* : *Call* $\rightarrow$ *Name* that returns the name of a call. |
| *match* | There is an operation *match* : *Call* $\rightarrow$ *Cond* that, given a call expression, e.g, $f(a\ :\ [])$, returns a condition expression that returns true if that expression matches a concrete call. For instance, in the concrete syntax example in rule 11, this yields the expression matchPattern(inp1, a:[]), assuming a method matchPattern exists. |
| *callExpr* | A function *callExpr* : *Call* $\rightarrow$ *Call* returns a call expression for a given method call replacing parameter expressions by simple names. For example, for a method f([], x:[]), the operation might return f(inp1, inp2). Note, that we assume that operations *callExpr* and *match* consistently name input parameters, e.g., inp1, inp2, . . .. |
| *isException* | A function *isException* : *Call* $\rightarrow \mathbb{B}$ determines if the call is a raised exception. |

**Helpers:**

| Name | Definition |
|---|---|
| *substates* | Given a state $s$ and a Statechart $sc$, the substates of s are given as $\mathit{substates}(s, sc) = \{s' \mid s' \in \mathit{sc.states} \wedge (s'.\mathit{stateName}, s.\mathit{stateName}) \in \mathit{sc.sub}\}$ |
| *superstates* | Given a state $s$ and a Statechart $sc$, the superstates of s are given as $\mathit{superstates}(s, sc) = \{s' \mid s' \in \mathit{sc.states} \wedge (s.\mathit{stateName}, s'.\mathit{stateName}) \in \mathit{sc.sub}^+\}$ |
| *ingoingT* | Given a state $s$ and a Statechart $sc$, the set of ingoing transitions to s is given as $\mathit{ingoingT}(s, sc) = \{t \mid t \in \mathit{sc.trans} \wedge t.trg = \mathit{s.stateName}\}$ |
| *outgoingT* | Given a state $s$ and a Statechart $sc$, the set of outgoing transitions from s is given as $\mathit{outgoingT}(s, sc) = \{t \mid t \in \mathit{sc.trans} \wedge t.src = \mathit{s.stateName}\}$ |
| *topInitials*, *topInitial* | $\mathit{topInitials}(sc)$ is the set of all top-level initial states: $\mathit{topInitials}(sc) = \{s \in \mathit{sc.states} \mid \mathsf{initial} \in \mathit{s.modifiers} \wedge \nexists s' \in \mathit{sc.states} : (\mathit{s.stateName}, s'.\mathit{stateName}) \in \mathit{sc.sub}\}$. If a Statechart $sc$ has an initial top-level state $\mathit{topInitial}(sc)$ is true: $\mathit{topInitial}(sc) = \mathsf{true} \iff \mathit{topInitials}(sc) \neq \emptyset$ |
| *topFinals*, *topFinal* | $\mathit{topFinals}(sc)$ is the set of all top-level final states: $\mathit{topFinals}(sc) = \{s \in \mathit{sc.states} \mid \mathsf{final} \in \mathit{s.modifiers} \wedge \nexists s' \in \mathit{sc.states} : (\mathit{s.stateName}, s'.\mathit{stateName}) \in \mathit{sc.sub}\}$. If a Statechart $sc$ has a final top-level state $\mathit{topFinal}(sc)$ is true: $\mathit{topFinal}(sc) = \mathsf{true} \iff \mathit{topFinals}(sc) \neq \emptyset$ |
| *simpleState* | A state $s$ of a Statechart $sc$ is a *simple* state if it has no substates and does not contain do actions or internal transitions that eventually may be transformed to substates: $\mathit{simpleState}(s, sc) = \mathsf{true} \iff \mathit{substates}(s, sc) = \emptyset \wedge \mathit{s.do} = \varepsilon \wedge \mathit{s.internT} = \emptyset$ |
| *initialIrrelevant* | Given a Statechart $sc$ and a state $s$ with modifier initial, $\mathit{initialIrrelevant}(s, sc)$ checks if that modifier is irrelevant (not needed to forward ingoing transitions or life cycle relevant): $\mathit{initialIrrelevant}(s, sc) = \mathsf{true} \iff s \notin \mathit{topInitials}(sc) \wedge \mathit{topInitial}(sc) \wedge \forall \mathit{sup} \in (\mathit{superstates}(s, sc) \cup \{s\}) : \mathit{ingoingT}(\mathit{sup}, sc) = \emptyset \wedge \nexists \mathit{sup}' \in \mathit{topInitials}(sc) : (\mathit{s.stateName}, \mathit{sup}'.\mathit{stateName}) \in \mathit{sc.sub}^+$ |
| *finalIrrelevant* | Given a Statechart $sc$ and a state $s$ with modifier final, $\mathit{finalIrrelevant}(s, sc)$ checks if that modifier is irrelevant (not needed for leading back outgoing transitions or life cylce relevant): $\mathit{finalIrrelevant}(s, sc) = \mathsf{true} \iff s \notin \mathit{topFinals}(sc) \wedge \mathit{topFinal}(sc) \wedge \forall \mathit{sup} \in (\mathit{superstates}(s, sc) \cup \{s\}) : \mathit{outgoingT}(\mathit{sup}, sc) = \emptyset \wedge \nexists \mathit{sup}' \in \mathit{topFinals}(sc) : (\mathit{s.stateName}, \mathit{sup}'.\mathit{stateName}) \in \mathit{sc.sub}^+$ |
| *sameCall* | Given a Statechart $sc$, a set of transitions $ts$ and a state $s$, $\mathit{sameCall}(s, ts, sc)$ checks if all calls have the same name and no other outgoing transition from s has that name. $\mathit{sameCall}(s, ts, sc) = \mathsf{true} \iff \forall t_1, t_2 \in ts : \mathit{nameOf}(t_1.\mathit{call}) = \mathit{nameOf}(t_2.\mathit{call}) \wedge \nexists t \in \mathit{outgoingT}(s, sc) \setminus ts : \mathit{sameCall}(s, (ts \cup \{t\}), sc)$ |
| *noPrio* | Given a transition $t$, $\mathit{noPrio}(t)$ checks if the transitions has no stereotype that defines a priority. $\mathit{noPrio}(t) = \mathsf{true} \iff \mathsf{prio = n} \notin \mathit{t.stereos}$ $(n \in \mathbb{N})$ |
| *noPrios* | Given a set of transitions $ts$, $\mathit{noPrios}(ts)$ checks if no transition has a stereotype that defines a priority. $\mathit{noPrios}(ts) = \mathsf{true} \iff \forall t \in ts : \mathit{noPrio}(t)$ |
| *listOfAllSuperstates*, *listOfSuperstates* | $\mathit{listOfAllSuperstates}(s, sc)$ returns all superstates of $s$ in a Statechart $sc$ as a list: $\mathit{listOfAllSuperstates}(s, sc) = [s_1, \ldots, s_n]$ where $(s_i.\mathit{stateName}, s_{i+1}.\mathit{stateName}) \in \mathit{sc.sub}, 1 \leq i < n$. The list of superstates of a state $s$ up to a state $s_t$ is $\mathit{listOfSuperstates}(s, s_t, sc) = [s_1, \ldots, s_n]$ where $(s_i.\mathit{stateName}, s_{i+1}.\mathit{stateName}) \in \mathit{sc.sub}, 1 \leq i < n$ and $s_n = s_t$. |
| *filter* | Function $\mathit{filter}(p, l)$ returns a list (or set) that only contains elements from $l$ for which predicate $p$ holds. |
| *cs* | The set of common superstates of states $s_1$ and $s_2$ in a Statechart $sc$ is defined as: $\mathit{cs}(s_1, s_2, sc) = \{s \in \mathit{sc.states} \mid (s_1.\mathit{stateName}, s.\mathit{stateName}) \in \mathit{sc.sub}^+ \wedge (s_2.\mathit{stateName}, s.\mathit{stateName}) \in \mathit{sc.sub}^+\}$ |
| *lcs* | The least common superstate hence is: $s = \mathit{lcs}(s_1, s_2, sc) \iff s \in \mathit{cs}(s_1, s_2, sc) \wedge \forall s' \in \mathit{cs}(s_1, s_2, sc) : (s.\mathit{stateName}, s'.\mathit{stateName}) \in \mathit{sc.sub}^+$ |
| *flatAndSimplified* | A Statechart $sc$ can be regarded as flat and simplified if the following condition holds: there are no do, entry, and exit actions, hierarchical states do not contain information, i.e., have no ingoing or outgoing transitions and no invariant, there are no superfluous initial or final states. $$\begin{array}{l} \mathit{flatAndSimplified}(sc) \iff \\ \forall s \in \mathit{sc.states} : \\ \quad s.\mathit{do}, s.\mathit{entry}, s.\mathit{exit} = \varepsilon \wedge \\ \quad s.\mathit{internT} = \emptyset \wedge \\ \quad \mathit{substates}(s, sc) \neq \emptyset \implies \\ \qquad (\mathit{ingoingT}(s, sc) = \emptyset \wedge \mathit{outgoingT}(s, sc) = \emptyset \wedge s.\mathit{inv} = \varepsilon) \wedge \\ \quad \mathit{initialIrrelevant}(s, sc) \implies \mathsf{initial} \notin s.\mathit{modifiers} \wedge \\ \quad \mathit{finalIrrelevant}(s, sc) \implies \mathsf{final} \notin s.\mathit{modifiers} \end{array}$$ |

## 5.3 Transformation Rules

| 1. elimDo | |
|---|---|
| bind | $sc \in SCFull$<br>$(stereos_s, modifier, n_s, inv_s, entry, exit, do, internT) \in sc.states$ |
| pre | $do \neq \varepsilon$ |
| $\Delta$ | $entry, exit, do, internT$ |
| trafo | $internT' = internT \cup \{(true, timeout, (do.stmt \;\&\; setTimer, do.cond))\}$<br>$entry' = (entry.stmt \;\&\; setTimer, entry.cond)$<br>$exit' = (exit.stmt \;\&\; stopTimer, exit.cond)$<br>$do' = \varepsilon$ |
| comment | (re)enables elimInternalT* |
| example | c.f. [11, p. 223, fig. 6.48] |

| 2. elimInternalT1 | |
|---|---|
| bind | $sc = (stereos_{sc}, n_{sc}, cl, inv_{sc}, states, trans, sub) \in SCFull$<br>$s = (stereos_s, modifier, n_s, inv_s, entry, exit, do, internT) \in states$<br>$inT = (pre, call, act) \in internT$ |
| pre | $substates(s, sc) \neq \emptyset$ |
| $\Delta$ | $internT, trans$ |
| trafo | $internT' = internT \backslash \{inT\}$<br>$trans' = trans \cup$<br>$(\bigcup_{s_i \in substates(s,sc)} \{(\emptyset, s_i.stateName, pre, call, act, s_i.stateName)\})$ |
| comment | (re)enables forwardToSub, backwardToSub* |
| example | 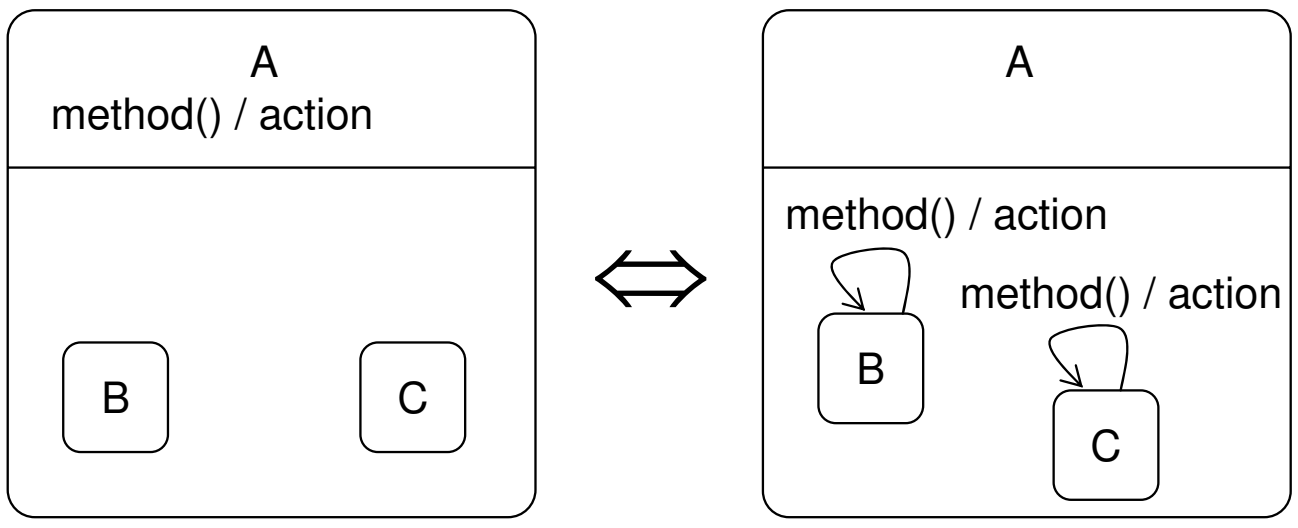 |

| 3. elimInternalT2 | |
|---|---|
| bind | $sc = (stereos_{sc}, n_{sc}, cl, inv_{sc}, states, trans, sub) \in SCFull$<br>$s = (stereos_s, modifier, n_s, inv_s, entry, exit, do, internT) \in states$<br>$inT = (pre, call, act) \in internT$ |
| pre | $substates(s, sc) = \emptyset$ |
| $\Delta$ | $internT, trans, states, sub$ |
| trafo | $internT' = internT \setminus \{inT\}$<br>$\lvert states' \setminus states \rvert = 1$<br>$\exists s' \in states' \setminus states :$<br>$(s'.stateName, s.stateName) \in sub' \wedge \{\mathsf{initial}, \mathsf{final}\} \subseteq s'.modifiers$<br>$trans' = trans \cup \{(\emptyset, s'.stateName, pre, call, act, s'.stateName)\}$ |
| comment | (re)enables forwardToSub, backwardToSub* |
| example | c.f. [11, p. 223, fig. 6.47] |

| 4. addInitTop | |
|---|---|
| bind | $sc \in SCFull$<br>$modifiers = \{s.modifiers \mid s \in sc.states \wedge superstates(s, sc) = \emptyset\}$<br>$\{modifiers_1, \ldots, modifiers_n\} = modifiers$ |
| pre | $n > 0 \wedge$<br>$\forall m \in modifiers : \mathsf{initial} \notin m$ |
| $\Delta$ | $modifiers_i, 1 \leq i \leq n$ |
| trafo | $modifiers'_i = modifiers_i \cup \{\mathsf{initial}\}$ |
| comment | If no top-level state is marked as initial, all top-level states are considered initial. |
| example | similar to [11, p. 202, fig. 6.27], (only consider top-level states) |

| 5. addInitSub | |
|---|---|
| bind | $sc \in SCFull$<br>$s \in sc.states$<br>$modifiers = \{st.modifiers \mid st \in substates(s, sc)\}$<br>$\{modifiers_1, \ldots, modifiers_n\} = modifiers$ |
| pre | $n > 0 \wedge$<br>$(\mathsf{initial} \in s.modifiers \vee ingoingT(s, sc) \neq \emptyset) \wedge$<br>$\forall m \in modifiers : \mathsf{initial} \notin m$ |
| $\Delta$ | $modifiers_i, 1 \leq i \leq n$ |
| trafo | $modifiers'_i = modifiers_i \cup \{\mathsf{initial}\}$ |
| comment | (re)enables forwardToSub, addInitSub. Only needed if parent state is initial or if transitions have to be forwarded. |
| example | c.f. [11, p. 202, fig. 6.27] |

| 6. forwardToSub | |
|---|---|
| bind | $sc \in SCFull$<br>$s \in sc.states$<br>$trans = sc.trans$<br>$subs = \{st \| st \in substates(s, sc) \wedge \textsf{initial} \in st.modifiers\}$<br>$t = (stereos, src, pre, call, act, trg) \in ingoingT(s.sc)$ |
| pre | $subs \neq \emptyset$ |
| $\Delta$ | $trans$ |
| trafo | $trans' = (trans \backslash \{t\}) \cup$<br>$(\bigcup_{sub \in subs} \{(stereos, src, pre, call, act, sub.stateName)\})$ |
| comment | (re)enables forwardToSub, addInitSub |
| example | c.f. [11, p. 202, fig. 6.25] |

| 7. deleteInitSub | |
|---|---|
| bind | $sc \in SCFull$<br>$s \in sc.states$<br>$modifiers = s.modifiers$ |
| pre | $\textsf{initial} \in modifiers \wedge$<br>$initialIrrelevant(s, sc)$ |
| $\Delta$ | $modifiers$ |
| trafo | $modifiers' = modifiers \backslash \{\textsf{initial}\}$ |
| comment | Irrelevant initial modifiers can be removed. In the example, initial modifiers of Z can be removed since no transitions need to be forwarded to Z and the top-level state A is not initial (not relevant for life cycle). Initial state of X cannot be deleted, otherwise X and Y would become life cycle relevant and transition f would be forwarded to both states. |
| example | 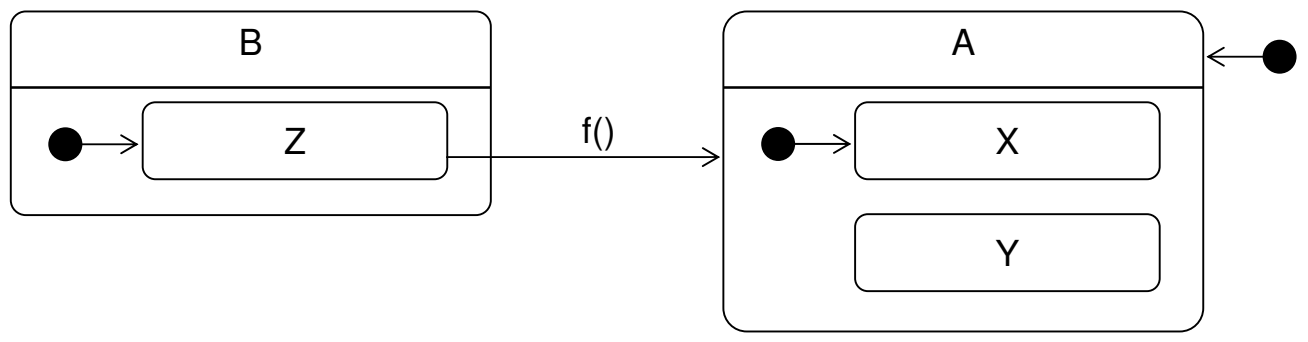  |

| 8. addFinalTop | |
|---|---|
| bind | $sc \in SCFull$<br>$modifiers = \{s.modifiers \| s \in sc.states \wedge superstates(s, sc) = \emptyset\}$<br>$\{modifiers_1, \ldots, modifiers_n\} = modifiers$ |
| pre | $n > 0 \wedge$<br>$\forall m \in modifiers : \mathsf{final} \notin m$ |
| Δ | $modifiers_i, 1 \leq i \leq n$ |
| trafo | $modifiers'_i = modifiers_i \cup \{\mathsf{final}\}$ |
| comment | If no top-level state is marked as final, all top-level states are considered final. |
| example | similar to [11, p. 202, fig. 6.27], (only consider top-level states) |

| 9. addFinalSub | |
|---|---|
| bind | $sc \in SCFull$<br>$s \in sc.states$<br>$modifiers = \{st.modifiers \| st \in substates(s, sc)\}$<br>$\{modifiers_1, \ldots, modifiers_n\} = modifiers$ |
| pre | $n > 0 \wedge$<br>$(\mathsf{final} \in s.modifiers \vee outgoingT(s, sc) \neq \emptyset) \wedge$<br>$\forall m \in modifiers : \mathsf{final} \notin m$ |
| Δ | $modifiers_i, 1 \leq i \leq n$ |
| trafo | $modifiers'_i = modifiers_i \cup \{\mathsf{final}\}$ |
| comment | (re)enables backwardToSub*, addFinalSub |
| example | c.f. [11, p. 202, fig. 6.27] |

| 10. backwardToSub | |
|---|---|
| bind | $sc \in SCFull$<br>$s \in sc.states$<br>$trans = sc.trans$<br>$subs = \{st \mid st \in substates(s, sc) \wedge \mathsf{final} \in st.modifiers\}$<br>$t = (stereos, src, pre, call, act, trg) \in outgoingT(s, sc)$ |
| pre | $\mathsf{prio{:}inner}, \mathsf{prio{:}outer} \notin sc.stereos \wedge$<br>$subs \neq \emptyset$ |
| $\Delta$ | $trans$ |
| trafo | $trans' = (trans \setminus \{t\}) \cup$<br>$(\bigcup_{sub \in subs} \{(stereos, sub.stateName, pre, call, act, trg)\})$ |
| comment | (re)enables backwardToSub. If no priority scheme for transitions is defined, transitions are led back unchanged to all final substates. Note: Also applied to transitions with manual priorities. Overlaps with rule 13 which leads back transitions with priorities but here all transitions are lead back (because no priority stereotype for the Statechart is defined). |
| example | c.f. [11, p. 202, fig. 6.26] |

| 11. backwardToSubPrioInner | |
|---|---|
| bind | $sc \in SCFull$<br>$s \in sc.states$<br>$trans = sc.trans$<br>$subs = \{st \mid st \in substates(s, sc) \wedge \mathsf{final} \in st.modifiers\}$<br>$\{s_1, \ldots, s_n\} = subs$<br>$ts \subseteq outgoingT(s, sc)$ |
| pre | $\mathsf{prio{:}inner} \in sc.stereos \wedge$<br>$subs \neq \emptyset \wedge$<br>$ts \neq \emptyset \wedge sameCall(s,ts,sc) \wedge noPrios(ts)$ |
| $\Delta$ | $trans$ |
| trafo | $cTrans_i = \{t \mid t \in outgoingT(s_i, sc) \wedge sameCall(s, \{t\} \cup ts, sc) \wedge noPrio(t)\}$<br>$preCond_i = \&\&_{t \in cTrans_i} !(t.pre\ \&\&\ match(t.call))$<br>$trans_i = \bigcup_{t \in ts}$<br>$\{(t.stereos, s_i.stateName, (t.pre\ \&\&\ preCond_i), callExpr(t.call), t.act, t.trg)\}$<br>$trans' = (trans \backslash ts) \cup (\bigcup_i trans_i)$ |
| comment | (re)enables backwardToSubPrioInner. Transitions with the same trigger name have to be led back at once and to all substates (marked as final) because their preconditions have to be adapted to reflect their priority (only if inner transitions' preconditions are not fulfilled or if their call expressions do not match, the outer transition is taken). |
| example | 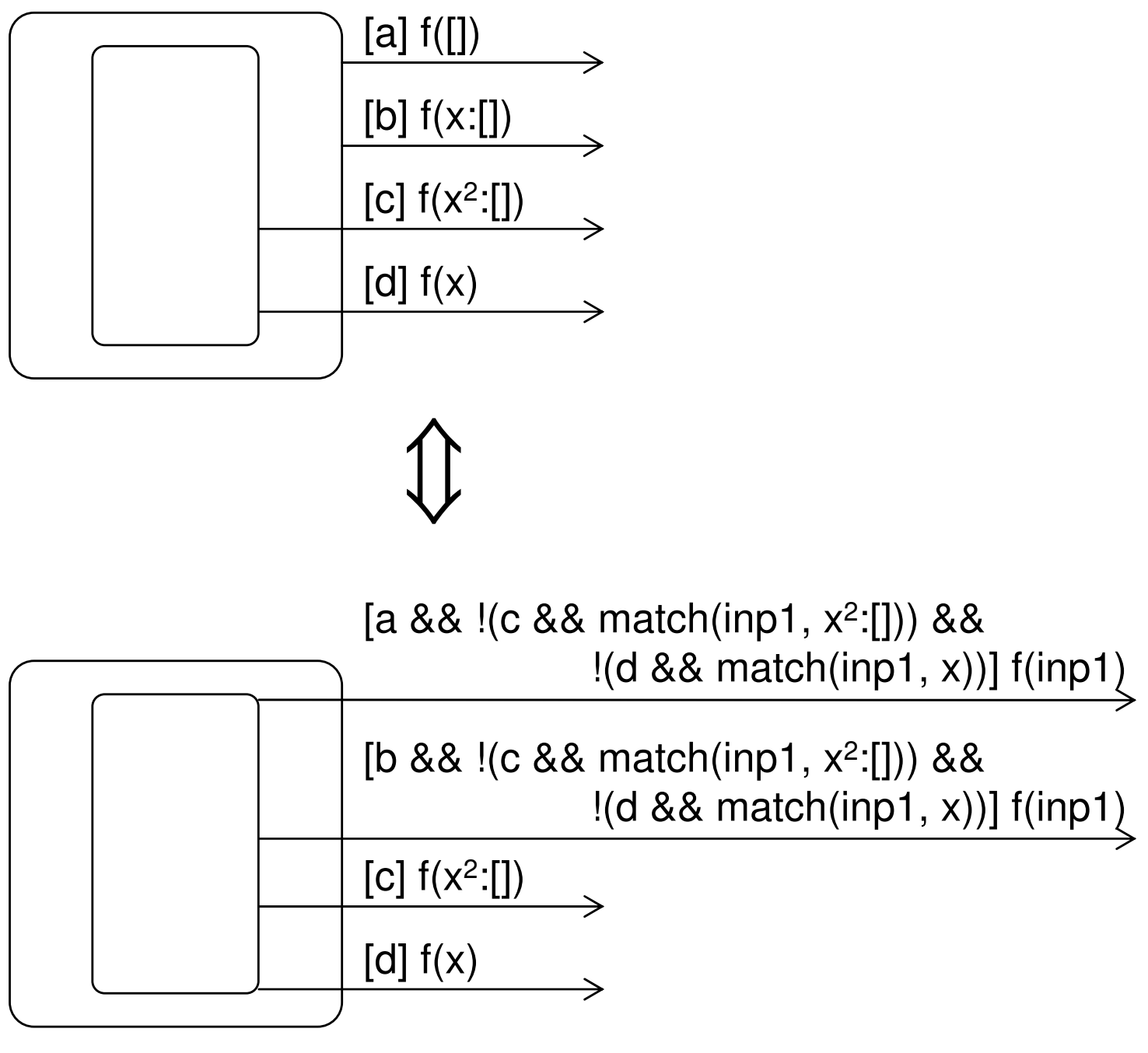 |

| 12. backwardToSubPrioOuter | |
|---|---|
| bind | $sc \in SCFull$<br>$s \in sc.states$<br>$trans = sc.trans$<br>$subs = \{st \mid st \in substates(s, sc) \wedge \textsf{final} \in st.modifiers\}$<br>$\{s_1, \ldots, s_n\} = subs$<br>$ts \subseteq outgoingT(s, sc)$ |
| pre | $\textsf{prio:outer} \in sc.stereos \wedge$<br>$subs \neq \emptyset \wedge$<br>$ts \neq \emptyset \wedge sameCall(s, ts, sc) \wedge noPrios(ts)$ |
| $\Delta$ | $trans$ |
| trafo | $preCond = \&\&_{t \in ts}!(t.pre \;\&\&\; match(t.call))$<br>$cTrans_i = \{t \mid t \in outgoingT(s_i, sc) \wedge sameCall(s, \{t\} \cup ts) \wedge noPrio(t)\}$<br>$trans_i = \bigcup_{t \in ts}\{(t.stereos, s_i.stateName, t.pre, t.call, t.act, t.trg)\}$<br>$trans_i^* =$<br>$\bigcup_{t \in cTrans_i}$<br>$\{(t.stereos, t.src, (t.pre \;\&\&\; preCond_i), callExpr(t.call), t.act, t.post, t.trg)\}$<br>$trans' = (trans \backslash ((\bigcup_i cTrans_i) \cup ts)) \cup (\bigcup_i trans_i) \cup (\bigcup_i trans_i^*)$ |
| comment | (re)enables backwardToSubPrioOuter.<br>$trans_i$ only adapts source state of higher priority outer transitions.<br>$trans_i^*$ leaves source and target alone but adapts precondition of lower priority inner transitions. |
| example | 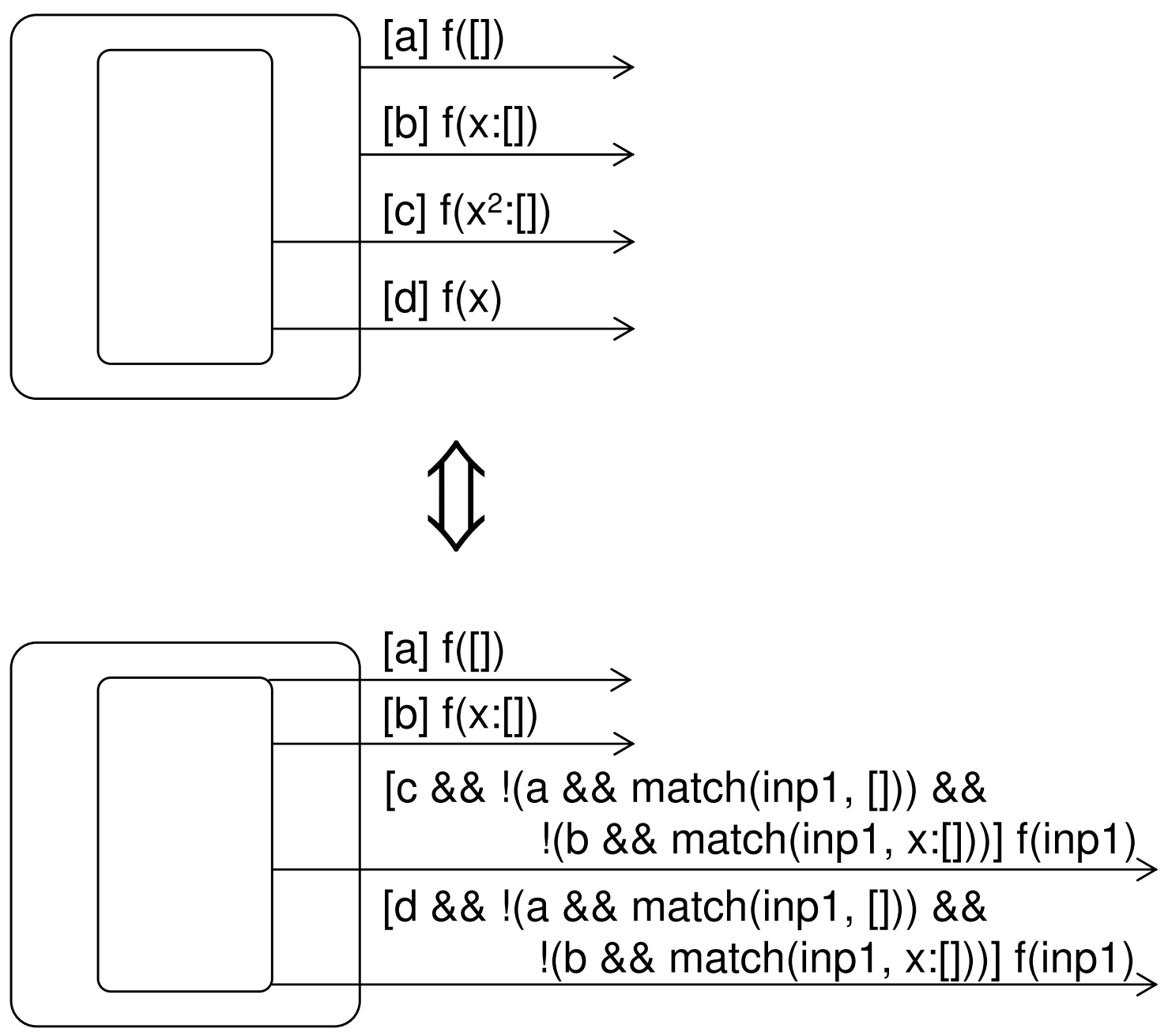 |

| 13. backwardToSubPrio | |
|---|---|
| bind | $sc \in SCFull$<br>$s \in sc.states$<br>$trans = sc.trans$<br>$subs = \{st \vert st \in substates(s) \wedge \textsf{final} \in st.modifiers\}$<br>$t = (stereos, src, pre, call, act, trg) \in outgoingT(s, sc)$ |
| pre | $\exists n \in \mathbb{N}$ : prio = n $\in stereos \wedge$<br>$subs \neq \emptyset$ |
| $\Delta$ | $trans$ |
| trafo | $trans' = (trans \backslash \{t\}) \cup (\bigcup_{sub \in subs} \{(stereos, sub.stateName, pre, call, act, trg)\})$ |
| comment | (re)enables backwardToSubPrio. Priority information is kept by this transformation, hence each transition may be led back separately. |
| example | similar to [11, p. 202, fig. 6.26], (lead back transitions with manual priorites unchanged) |

| 14. elimPrio | |
|---|---|
| bind | $sc \in \mathit{SCFull}$<br>$s \in \mathit{sc.states}$<br>$\mathit{trans} = \mathit{sc.trans}$<br>$\mathit{ts} \subseteq \mathit{outgoingT}(s, sc)$<br>$\{t_1, \ldots, t_n\} = \mathit{ts}$ |
| pre | $\mathit{simpleState}(s, sc) \land$<br>$\forall \mathit{sup} \in \mathit{superstates}(s, sc) : \mathit{outgoingT}(sup, sc) = \emptyset \land$<br>$\mathit{ts} \neq \emptyset \land \mathit{sameCall}(s, ts, sc) \land (\forall t \in ts : \neg \mathit{noPrios}(ts))$ |
| $\Delta$ | $\{t_1, \ldots, t_n\}$ |
| trafo | $ts_{h,i} = \{t \mid t \in ts \land \mathit{prio}(t) > \mathit{prio}(t_i)\}$<br>$\mathit{precond}_i = \&\&_{t \in ts_{h,i}} !(t.\mathit{pre}\ \&\&\ \mathit{match}(t.\mathit{call}))$<br>$t'_i = (\emptyset, t_i.\mathit{src}, (t_i.\mathit{pre}\ \&\&\ \mathit{precond}_i), \mathit{callExpr}(t_i.\mathit{call}), t_i.\mathit{act}, t_i.\mathit{trg})$ |
| comment | May only convert explicit priorities into preconditions if all superstates do not have transitions to be led back. |
| example | 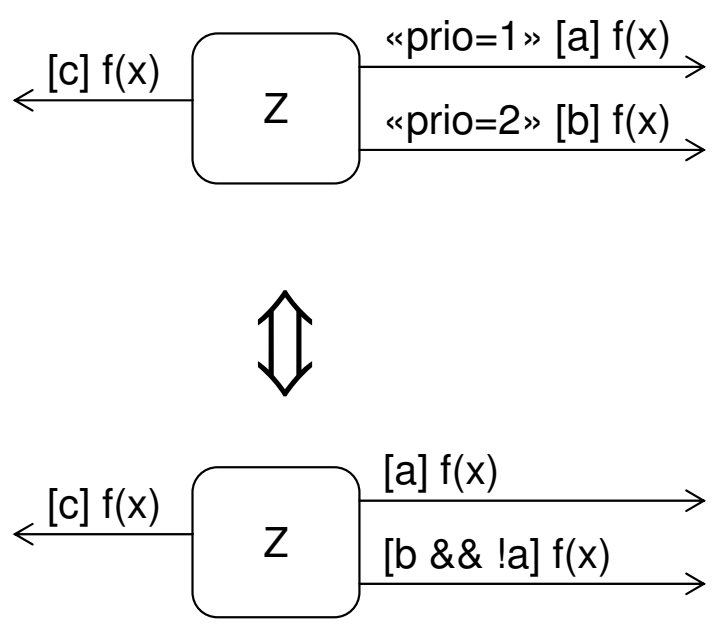 |

| 15. deleteFinalSub | |
|---|---|
| bind | $sc \in SCFull$<br>$s \in sc.states$<br>$modifiers = s.modifiers$ |
| pre | $\mathsf{final} \in modifiers$<br>$finalIrrelevant(s, sc)$ |
| $\Delta$ | $modifiers$ |
| trafo | $modifiers' = modifiers \backslash \{\mathsf{final}\}$ |
| comment | Irrelevant final modifiers can be removed. In the example, final modifiers of Z can be deleted since no transition needs to be led back from B to Z and top-level state B is not relevant for life cycle. Final stereotype of X cannot be removed, otherwise X and Y would become relevant for life cycle and transition f would be led back to both states. |
| example | 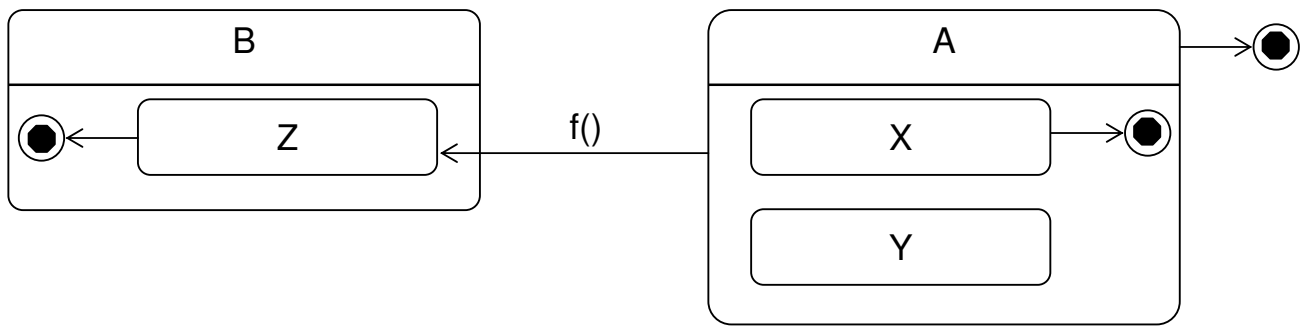 |

| 16. moveExitActions | |
|---|---|
| bind | $sc \in SCFull$<br>$s \in sc.states$<br>$exit = s.exit$<br>$act = exit.stmt$<br>$post = exit.cond$<br>$\{t_1, \ldots, t_n\} = outgoingT(s, sc)$ |
| pre | $exit \neq \varepsilon \wedge$<br>$\mathsf{final} \notin s.modifiers \wedge$<br>$simpleState(s) \wedge$<br>$\forall sup \in superstates(s, sc) : (outgoingT(sup, sc) = \emptyset) \wedge$<br>$\mathsf{action\ conditions{:}sequential} \notin sc.stereos$ |
| $\Delta$ | $exit, \{t_1, \ldots, t_n\}$ |
| trafo | $exit' = \varepsilon$<br>$[s_1, \ldots, s_n] = filter(listOfSuperstates(s, lcs(s, t_i.trg, sc), sc), (\lambda x\ .x.exit \neq \varepsilon))$<br>$act_i = act \& s_1.exit.stmt \& \ldots \& s_n.exit.stmt \& t_i.act.stmt$<br>$post_i = post\ \&\&\ s_1.exit.cond\ \&\&\ \ldots\ \&\&\ s_n.exit.cond\ \&\&\ t_i.act.cond$<br>$t_i' = (t_i.stereos, t_i.src, t_i.pre, t_i.call, (act_i, post_i), t_i.trg)$ |
| comment | Exit actions of all superstates up to the least common superstate are taken into account. Has to be done for all outgoing transitions because we cannot, in general, decide if a transition already contains the exit actions. |
| example | c.f. [11, p. 220, fig. 6.43,6.44], (only exit part) |

| 17. moveExitActionsSeq | |
|---|---|
| bind | $sc \in SCFull$<br>$s \in sc.states$<br>$exit = s.exit$<br>$\{t_1, \ldots, t_n\} = outgoingT(s, sc)$ |
| pre | $exit \neq \varepsilon \wedge$<br>final $\notin s.modifiers \wedge$<br>$simpleState(s) \wedge$<br>$\forall sup \in superstates(s, sc) : (outgoingT(sup,sc) = \emptyset) \wedge$<br>action conditions:sequential $\in sc.stereos$ |
| $\Delta$ | $exit, \{t_1, \ldots, t_n\}$ |
| trafo | $exit' = \varepsilon$<br>$[s_1, \ldots, s_n] = filter(listOfSuperstates(s, lcs(s, t_i.trg, sc), sc), (\lambda x\ .x.exit \neq \varepsilon))$<br>$seq_i = exit + s_1.exit + \ldots + s_n.exit + t_i.act$<br>$act_i = (seq_i.stmt, t_i.act.cond)$<br>$t'_i = (t_i.stereos, t_i.src, t_i.pre, t_i.call, act_i, t_i.trg)$ |
| comment | Similar to moveExitActions but conditions are transformed to sequentially valid actions. |
| example | c.f. [11, p. 221, fig. 6.45], (only exit part) |

| 18. removeExitAction | |
|---|---|
| bind | $sc \in SCFull$<br>$s \in sc.states$<br>$exit = s.exit$ |
| pre | $exit \neq \varepsilon \wedge$<br>$outgoingT(s, sc) = \emptyset \wedge$<br>$\forall sup \in superstates(s, sc) : (outgoingT(sup,sc) = \emptyset)$ |
| $\Delta$ | $exit$ |
| trafo | $exit' = \varepsilon$ |
| comment | Exit actions if states that can never be left can be removed. |

| 19. moveEntryActions | |
|---|---|
| bind | $sc \in SCFull$<br>$s \in sc.states$<br>$entry = s.entry$<br>$act = entry.stmt$<br>$post = entry.cond$<br>$\{t_1, \ldots, t_n\} = ingoingT(s, sc)$ |
| pre | $entry \neq \varepsilon \wedge$<br>initial $\notin s.modifiers \wedge$<br>$simpleState(s, sc) \wedge$<br>$\forall sup \in superstates(s, sc) : (ingoingT(sup, sc) = \emptyset) \wedge$<br>action conditions:sequential $\notin sc.stereos$ |
| $\Delta$ | $entry, \{t_1, \ldots, t_n\}$ |
| trafo | $entry' = \varepsilon$<br>$[s_1, \ldots, s_n] = filter(listOfSuperstates(s, lcs(s, t_i.src, sc), sc), (\lambda x.\ x.entry \neq \varepsilon))$<br>$act_i = t_i.act.stmt \& s_n.entry.stmt \& \ldots \& s_1.entry.stmt \& act$<br>$post_i = t_i.act.cond\ \&\&\ s_n.entry.cond\ \&\&\ \ldots\ \&\&\ s_1.entry.cond\ \&\&\ post$<br>$t'_i = (t_i.stereos, t_i.src, t_i.pre, t_i.call, (act_i, post_i), t_i.trg)$ |
| comment | Entry actions of all superstates up to the least common superstate are taken into account. Has to be done for all ingoing transitions because, in general, we cannot decide if a transition already contains the entry actions. |
| example | c.f. [11, p. 220, fig. 6.43,6.44], (only entry part) |

| 20. moveEntryActionsSeq | |
|---|---|
| bind | $sc \in \mathit{SCFull}$<br>$s \in sc.\mathit{states}$<br>$\mathit{entry} = s.\mathit{entry}$<br>$\mathit{post} = \mathit{entry}.\mathit{cond}$<br>$\{t_1, \ldots, t_n\} = \mathit{ingoingT}(s, sc)$ |
| pre | $\mathit{entry} \neq \varepsilon \wedge$<br>initial $\notin s.\mathit{modifiers} \wedge$<br>$\mathit{simpleState}(s, sc) \wedge$<br>$\forall \mathit{sup} \in \mathit{superstates}(s, sc) : (\mathit{ingoingT}(\mathit{sup}, sc) = \emptyset) \wedge$<br>action conditions:sequential $\in sc.\mathit{stereos}$ |
| $\Delta$ | $\mathit{entry}, \{t_1, \ldots, t_n\}$ |
| trafo | $\mathit{entry}' = \varepsilon$<br>$[s_1, \ldots, s_n] = \mathit{filter}(\mathit{listOfSuperstates}(s, \mathit{lcs}(s, t_i.\mathit{src}, sc), sc), (\lambda x.\ x.\mathit{entry} \neq \varepsilon))$<br>$\mathit{seq}_i = t_i.\mathit{act} + s_n.\mathit{entry} + \ldots + s_1.\mathit{entry} + \mathit{entry}$<br>$\mathit{act}_i = (\mathit{seq}_i.\mathit{stmt}, t_i.\mathit{act}.\mathit{cond}\ \&\&\ \mathit{entry}.\mathit{cond})$<br>$t_i' = (t_i.\mathit{stereos}, t_i.\mathit{src}, t_i.\mathit{pre}, t_i.\mathit{call}, \mathit{act}_i, t_i.\mathit{trg})$ |
| comment | Similar to moveEntryActions but conditions are transformed to sequentially valid actions. For the post condition, the transitions post condition and the entry actions's post condition must hold. |
| example | c.f. [11, p. 221, fig. 6.45], (only entry part) |

| 21. removeEntryAction | |
|---|---|
| bind | $sc \in \mathit{SCFull}$<br>$s \in sc.\mathit{states}$<br>$\mathit{entry} = s.\mathit{entry}$ |
| pre | $\mathit{entry} \neq \varepsilon \wedge$<br>$\mathit{ingoingT}(\mathit{ssc}) = \emptyset \wedge$<br>$\forall \mathit{sup} \in \mathit{superstates}(s, sc) : (\mathit{ingoingT}(\mathit{sup}, sc) = \emptyset)$ |
| $\Delta$ | $\mathit{entry}$ |
| trafo | $\mathit{entry}' = \varepsilon$ |
| comment | Entry actions of states that can never be entered can be removed. |

| 22. moveInvariant | |
|---|---|
| bind | $sc \in SCFull$<br>$states = sc.states$<br>$s \in states$<br>$subs = substates(s, sc)$<br>$inv = s.inv$ |
| pre | $inv \neq \varepsilon \wedge$<br>$subs \neq \emptyset \wedge$<br>$s.do = \varepsilon \wedge s.internT = \emptyset$ |
| $\Delta$ | $states, inv$ |
| trafo | $inv' = \varepsilon$<br>$states' = (states \backslash subs) \cup$<br>$(\bigcup_{sub \in subs} \{(sub.stereos, sub.modifier, sub.name, inv \&\& sub.inv,$<br>$sub.entry, sub.exit, sub.do, sub.internT)\}$ |
| comment | Done for a state (that will not be further transformed to a state with new substates) and all its substates since we cannot, in general, decide if a condition is already integrated in one substate. |
| example | c.f. [11, p. 197, fig. 6.18] |

| 23. removeHierarchy | |
|---|---|
| bind | $sc \in SCFull$<br>$states = sc.states$<br>$sub = sc.sub$<br>$superstates = \{s \in states \mid \exists s_{sub} \in states : (s_{sub}.stateName, s.stateName) \in sub\}$ |
| pre | $flatAndSimplified(sc) \wedge$<br>$sub \neq \emptyset$ |
| $\Delta$ | $states, sub$ |
| trafo | $sub' = \emptyset$<br>$states' = states \backslash superstates$ |
| comment | If Statechart is flat and simplified, superstates and the substate relation can be removed. |

| 24. completionIgnore | |
|---|---|
| bind | $sc \in \mathit{SCFull}$<br>$\mathit{trans} = sc.\mathit{trans}$<br>$\mathit{allCalls} = \{t.\mathit{call} \mid t \in \mathit{trans} \land \neg \mathit{isException}(t)\}$<br>$\{s_1, \ldots, s_n\} = sc.\mathit{states}$<br>$\mathit{outgoing}_i = \{t.\mathit{call} \mid t \in \mathit{outgoingT}(s_i, sc)\}, 1 \leq i \leq n$<br>$\mathit{missing}_i = \mathit{allCalls} \setminus \mathit{outgoing}_i, 1 \leq i \leq n$<br>$\{ts_i^1, \ldots, ts_i^{m_i}\} \subseteq \wp(\mathit{trans}), 1 \leq i \leq n$<br>$\mathit{call}_i^j = t.\mathit{call},$ where $t \in ts_i^j, 1 \leq i \leq n, 1 \leq j \leq m_i$ |
| pre | $\mathit{flatAndSimplified}(sc) \land$<br>$sc.\mathit{sub} = \emptyset \land$<br>completion:ignore $\in sc.\mathit{stereos} \land$<br>$ts_i^1 \cup \ldots \cup ts_i^{m_i} = \mathit{outgoingT}(s_i, sc), 1 \leq i \leq n \land$<br>$\mathit{sameCall}(s_i, ts_i^j), 1 \leq i \leq n, 1 \leq j \leq m_i$ |
| $\Delta$ | $\mathit{trans}$ |
| trafo | $\mathit{preCond}_i^j = \&\&_{t \in ts_i^j} !(t.\mathit{pre}\ \&\&\ \mathit{match}(t.\mathit{call}))$<br>$\mathit{compT}_i^j = \{(\emptyset, s_i.\mathit{stateName}, \mathit{preCond}_i^j, \mathit{callExpr}(\mathit{call}_i^j), \varepsilon, s_i.\mathit{stateName})\}$<br>$\mathit{compT}_i = \bigcup_{1 \leq j \leq m_i} \mathit{compT}_i^j$<br>$\mathit{newT}_i = \bigcup_{n \in \mathit{missing}_i} \{(\emptyset, s_i.\mathit{stateName}, \mathit{true}, \mathit{callExpr}(n), \varepsilon, s_i.\mathit{stateName})\}$<br>$\mathit{trans}_i = \mathit{compT}_i \cup \mathit{newT}_i$<br>$\mathit{trans}' = \mathit{trans} \cup (\bigcup_{1 \leq i \leq n} \mathit{trans}_i)$ |
| comment | $\mathit{newT}_i$ is the set of new transitions added for each missing transition, $\mathit{compT}_i$ is the set of transitions added with “negated” preconditions and call expressions so there is always an enabled transtion for each trigger. |
| example | 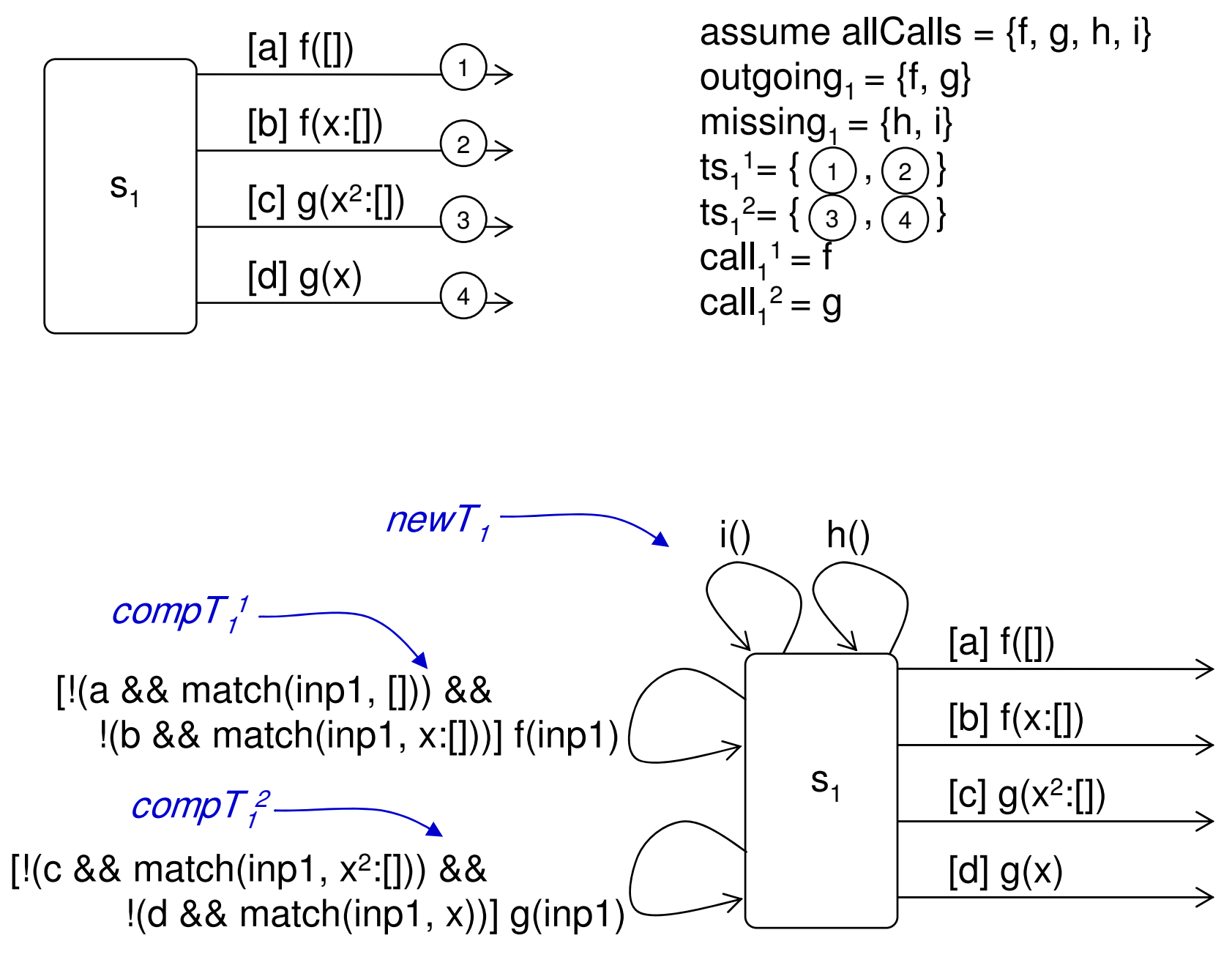 |

| 25. completionError | |
|---|---|
| bind | $sc \in \mathit{SCFull}$<br>$\mathit{trans} = sc.\mathit{trans}$<br>$\mathit{allCalls} = \{t.\mathit{call} \mid t \in \mathit{trans} \wedge \neg \mathit{isException}(t)\}$<br>$\{s_1, \ldots, s_n\} = sc.\mathit{states}$<br>$\mathit{outgoing}_i = \{t.\mathit{call} \mid t \in \mathit{outgoingT}(s_i, sc)\}, 1 \leq i \leq n$<br>$\mathit{missing}_i = \mathit{allCalls} \setminus \mathit{outgoing}_i, 1 \leq i \leq n$<br>$\{ts_i^1, \ldots, ts_i^{m_i}\} \subseteq \wp(\mathit{trans}), 1 \leq i \leq n$<br>$\mathit{call}_i^j = t.\mathit{call}$, where $t \in ts_i^j, 1 \leq i \leq n, 1 \leq j \leq m_i$<br>$s_{err} \in sc.\mathit{states}$ |
| pre | $\mathit{flatAndSimplified}(sc) \wedge$<br>error $\in s_{err}.\mathit{stereos} \wedge$<br>$sc.sub = \emptyset \wedge$<br>completion:error $\in sc.\mathit{stereos} \wedge$<br>$ts_i^1 \cup \ldots \cup ts_i^{m_i} = \mathit{outgoingT}(s_i, sc), 1 \leq i \leq n \wedge$<br>$\mathit{sameCall}(s_i, ts_i^j), 1 \leq i \leq n, 1 \leq j \leq m_i$ |
| $\Delta$ | $\mathit{trans}$ |
| trafo | $\mathit{preCond}_i^j = \&\&_{t \in ts_i^j} !(t.\mathit{pre}\ \&\&\ \mathit{match}(t.\mathit{call}))$<br>$\mathit{compT}_i^j = \{(\emptyset, s_i.\mathit{stateName}, \mathit{preCond}_i^j, \mathit{callExpr}(\mathit{call}_j^i), \varepsilon, s_{err}.\mathit{stateName})\}$<br>$\mathit{compT}_i = \bigcup_{1 \leq j \leq m_i} \mathit{compT}_i^j$<br>$\mathit{newT}_i = \bigcup_{n \in \mathit{missing}_i} \{(\emptyset, s_i.\mathit{stateName}, true, \mathit{callExpr}(n), \varepsilon, s_{err}.\mathit{stateName})\}$<br>$\mathit{trans}_i = \mathit{compT}_i \cup \mathit{newT}_i$<br>$\mathit{trans}' = \mathit{trans} \cup (\bigcup_{1 \leq i \leq n} trans_i)$ |
| comment | Same as completionIgnore but target of new transition is an error state (which is also completed). |

| 26. completionException | |
|---|---|
| bind | $sc \in SCFull$<br>$trans = sc.trans$<br>$allCalls = \{t.call \mid t \in trans \wedge isException(t)\}$<br>$\{s_1, \ldots, s_n\} = sc.states$<br>$outgoing_i = \{t.call \mid t \in outgoingT(s_i, sc)\}, 1 \leq i \leq n$<br>$missing_i = allCalls \backslash outgoing_i, 1 \leq i \leq n$<br>$\{ts_i^1, \ldots, ts_i^{m_i}\} \subseteq \wp(trans), 1 \leq i \leq n$<br>$call_i^j = t.call$, where $t \in ts_i^j, 1 \leq i \leq n, 1 \leq j \leq m_i$<br>$s_{ex} \in sc.states$ |
| pre | $flatAndSimplified(sc) \wedge$<br>exception $\in s_{ex}.stereos \wedge$<br>$sc.sub = \emptyset \wedge$<br>$ts_i^1 \cup \ldots \cup ts_i^{m_i} = outgoingT(s_i, sc), 1 \leq i \leq n \wedge$<br>$sameCall(s_i, ts_i^j), 1 \leq i \leq n, 1 \leq j \leq m_i$ |
| $\Delta$ | $trans$ |
| trafo | $preCond_i^j = \&\&_{t \in ts_i^j} !(t.pre \; \&\& \; match(t.call))$<br>$compT_i^j = \{(\emptyset, s_i.stateName, preCond_i^j, callExpr(call_j^i), \varepsilon, s_{ex}.stateName)\}$<br>$compT_i = \bigcup_{1 \leq j \leq m_i} compT_i^j$<br>$newT_i = \bigcup_{n \in missing_i} \{(\emptyset, s_i.stateName, true, callExpr(n), \varepsilon, s_{ex}.stateName)\}$<br>$trans_i = compT_i \cup newT_i$<br>$trans' = trans \cup (\bigcup_{1 \leq i \leq n} trans_i)$ |
| comment | Same as completionIgnore but target of new transition (with exception as trigger) is an exception state (which is also completed). |

There is no rule for chaos completion of Statechart as this is the default behavior.

Also note, that a considerable simplification of rules that involve adapting the preconditions can be achieved if the call expression would not need to be handled each time (by putting a match expression in the precondition). This would be in fact possible by starting with a transformation that replaces a complex call expressions by simple parameter names and adapts the precondition accordingly (as it is now done in every rule that adapts preconditions).

The relation *transformed* : $SCFull \times SCFull$ states that a Statechart is the transformation result of another after applying one transformation rule from above:

$$(sc, sc') \in transformed \Leftrightarrow$$
$$(sc, sc') \in (elimDo \cup \ldots \cup completionException)$$

A Statechart $sc' \in SCFull$ is the final transformation result of transforming a Statechart $sc$ if $(sc, sc') \in transformed^+ \wedge \nexists sc'' \in SCFull : (sc', sc'') \in transformed^+$.

In a final transformation result, the following constructs of Full UML/P Statecharts are not needed anymore:

- do actions,
- entry actions,
- exit actions,
- internal transitions,
- hierarchical states and sub state information,
- stereotypes.

# 6 Abstract Syntax of Simplified Statecharts

The abstract syntax of UML/P Statecharts can now be simplfied in order to avoid complex syntax expression together with constaints in the semantics mapping.

$$
\begin{array}{lcl}
\mathit{SCSimp} & = & \mathit{DiagramName} \times \mathit{ClassName} \times \mathit{Inv} \times \wp(\mathit{State}) \times \wp(\mathit{Transition}) \\
\mathit{State} & = & \wp(\mathit{Modifier}) \times \mathit{StateName} \times \mathit{Inv} \\
\mathit{Modifier} & = & \{\mathsf{initial}, \mathsf{final}\} \\
\mathit{Transition} & = & \mathit{Src} \times \mathit{Pre} \times \mathit{Call} \times \mathit{Act} \times \mathit{Trg} \\
\mathit{Inv}, \mathit{Pre} & = & \mathit{Cond} \\
\mathit{Act} & = & \mathit{Stmt} \times \mathit{Cond} \\
\mathit{Src}, \mathit{Trg} & = & \mathit{State} \\
\mathit{DiagramName}, & & \\
\mathit{ClassName}, & & \\
\mathit{StateName} & = & \mathit{Name}
\end{array}
$$

Note, that we have expanded the states names in transitions to the full states. The unspecified sets *Name*, *Cond*, and *Stmt* are the same as in Section 3.

If context conditions from above that are still applicable hold, the simplified Statechart is well-formed.

# 7 Mapping of Simplified UML/P Statecharts

A Statechart models the behavior of an object that is an instance of the class C (or any subclass) mentioned in the Statechart. The behavior is realized by one object or a group of objects. Groups of objects are for example needed to describe implementations that use the state pattern [7]. In any case, we assume a main object of class C.

First, we give some definitions that are helpful in addition to the system model definitions. They allow for a more concise semantics mapping later on.

## 7.1 Preliminaries

We introduce the name OGS (object group state) for the relation that describes the possible states of a set (or group) of objects:

$$\begin{aligned}\mathsf{OGS} = \mathsf{UOID} \rightarrow \quad & ((\mathsf{UVAR} \rightarrow \mathsf{UVAL})\times \\ & (\mathsf{UTHREAD} \rightarrow \mathit{Stack}(\mathsf{UFRAME}))\times \\ & \mathit{Buffer}(\mathsf{UEVENT}))\end{aligned}$$

The set of possible states of a set of objects *oids* is denoted $\mathit{states}(\mathit{oids})$ which is a subset of OGS. For each object identifier in *oids*, it is possible to extract the individual object's state from a state of a group of objects in $\mathit{states}(\mathit{oids})$.

We also define a function *buffer* to access an object's event buffer given the object identifier $o$ and a state of a group of objects *ogs*.

$$\begin{aligned}&\mathit{buffer} : \mathsf{OGS} \times \mathsf{UOID} \rightarrow \mathit{Buffer}(\mathsf{UEVENT}) \\ &\mathit{buffer}(\mathit{ogs}, o) = \pi_3(\mathit{ogs}(o))\end{aligned}$$

Given a stack of frames $s$, we also assume a function $\mathsf{proc}(s, m)$ that checks if the message $m$ is currently processed, i.e., if it's on top of the stack.

A valuation of type $V = \mathit{Name} \rightarrow \mathsf{UVAL}$ is a function that maps a name to a value. For the evaluation of an expression and the effect of a statement specified in the Statechart as invariant, pre- and postcondition, call trigger, or action, we assume the relations $\mathcal{C}$, $\mathcal{E}$, $\mathcal{S}$. We do not define these relations here in detail since we assume to import them just as we have done with their abstract syntax in Chapter 3. The valuation $V$ in that context may define variable bindings that cannot be extracted from the state information (e. g., transition preconditions that refer to the argument of a method. A valuation has to ensure that the actual value of the arguments can be obtained). $V$ also may not contradict the state information, i. e., define variable bindings that are already defined and bound to other values in the states. We assume relations instead of functions to enable underspecification also for expressions or statements.

- Boolean expressions

$$\mathcal{C} \subseteq \textit{Cond} \times \mathsf{OGS} \times \mathsf{UOID} \times \mathsf{UTHREAD} \times V$$

Given a state of objects *ogs*, an object identifier $o$, a thread $t$ , and a valuation $v$, $\mathcal{C}(\textit{cond}, \textit{ogs}, o, t, v)$ holds if the condition *cond* can be evaluated to true.

- Calls

$$\mathcal{E} \subseteq \textit{Call} \times \mathsf{OGS} \times \mathsf{UOID} \times \mathsf{UTHREAD} \times V \times V$$

$\mathcal{E}(\textit{call}, \textit{ogs}, o, t, v, v')$ holds if the (resulting) valuation $v'$ is an extension of the first such that all variables in *call* are bound in $v'$. For example, if the call is $f(a : as)$, the resulting valuation contains a variable binding for $a$ and $as$.

- Statements

$$\begin{aligned}\mathcal{S} \subseteq\ & \textit{Stmt} \times \mathsf{OGS} \times \mathsf{UOID} \times \mathsf{UTHREAD} \times V \times \\ & \mathsf{OGS} \times \mathsf{UMESSAGE}^*\end{aligned}$$

A statement may be composed of other statements, i. e., it may be a block statement. $\mathcal{S}(\textit{stmt}, \textit{ogs}, o, t, v, \textit{ogs}', \textit{ms})$ holds if the effect of the statement *stmt* is described by the differences between *ogs* and *ogs*$'$ and the messages sent.

To decide if a transition is enabled, among other things, there has to be a matching event in the object's event buffer. To find such an event that corresponds to a given call expression in the Statechart, we define a function msg.

$$\begin{aligned}&\mathsf{msg} : \textit{Call} \times \mathsf{Buffer(UEVENT)} \times \mathsf{UOID} \rightarrow \wp(\mathsf{UMESSAGE})\\ &\mathsf{msg}(f(p_1, \dots, p_k), \textit{buf}, o) = \{m \in \mathsf{callsOf}(o) | \mathsf{ReceiveEvent}(m) \in \textit{buf}\\ &\qquad \wedge \mathsf{opName}(m) = f \wedge \textit{length}(\mathsf{params}(m)) = k\}\end{aligned}$$

Since we do not have type information in the call expression, the selection of events from the buffer may be ambiguous. E. g., if a class provides two operations $f(\textit{bool } i)$ and $f(\textit{int } i)$ and the call expression is $f(i)$, we can't decide which operation is intended and possibly return messages for both.

Taking a Statechart transition must be somehow reflected in transitions of the group of objects realising the Statechart. Given two groups of object states $\textit{ogs}_s$ and $\textit{ogs}_t$, we are thus interested in checking if that second state can be reached from the first state by $n$ transition steps in $\Delta(\textit{oids})$ (see Fig. 7.1 that illustrates this relationship). Based on the definitions in [3], we define:

$$\begin{aligned}&\textit{ogs}_t \in \pi_1(\Delta(\textit{oids})^n(\textit{ogs}_s, _)) \iff\\ &\quad \exists g_1, \dots, g_n \in \textit{states}(\textit{oids}) :\\ &\quad g_1 \in \pi_1(\Delta(\textit{oids})(\textit{ogs}_s, _)) \wedge\\ &\quad g_{i+1} \in \pi_1(\Delta(\textit{oids})(g_i, _)) \wedge\\ &\quad \textit{ogs}_t \in \pi_1(\Delta(\textit{oids})(g_n, _))\end{aligned}$$

Finally, we assume a function $\textit{triggers} : \textit{SC} \rightarrow \wp(\textit{Name})$ that returns all trigger names, i.e., all state-relevant method names.

## 7.2 Mapping

A Statechart specifies the behavior of an object that may control additional objects in order to realize the required behavior. The general idea is to map Statechart states to states of a group of objects and formulate conditions that hold if the objects show a behavior conforming to the Statechart specification. Executing (or firing) a transition in the Statechart at point $t_0$ leads to a series of transitions in the system model's timed state transition system as depicted in Fig. 7.1, finally reaching a system model state that corresponds to a Statechart state at point $t_1$.

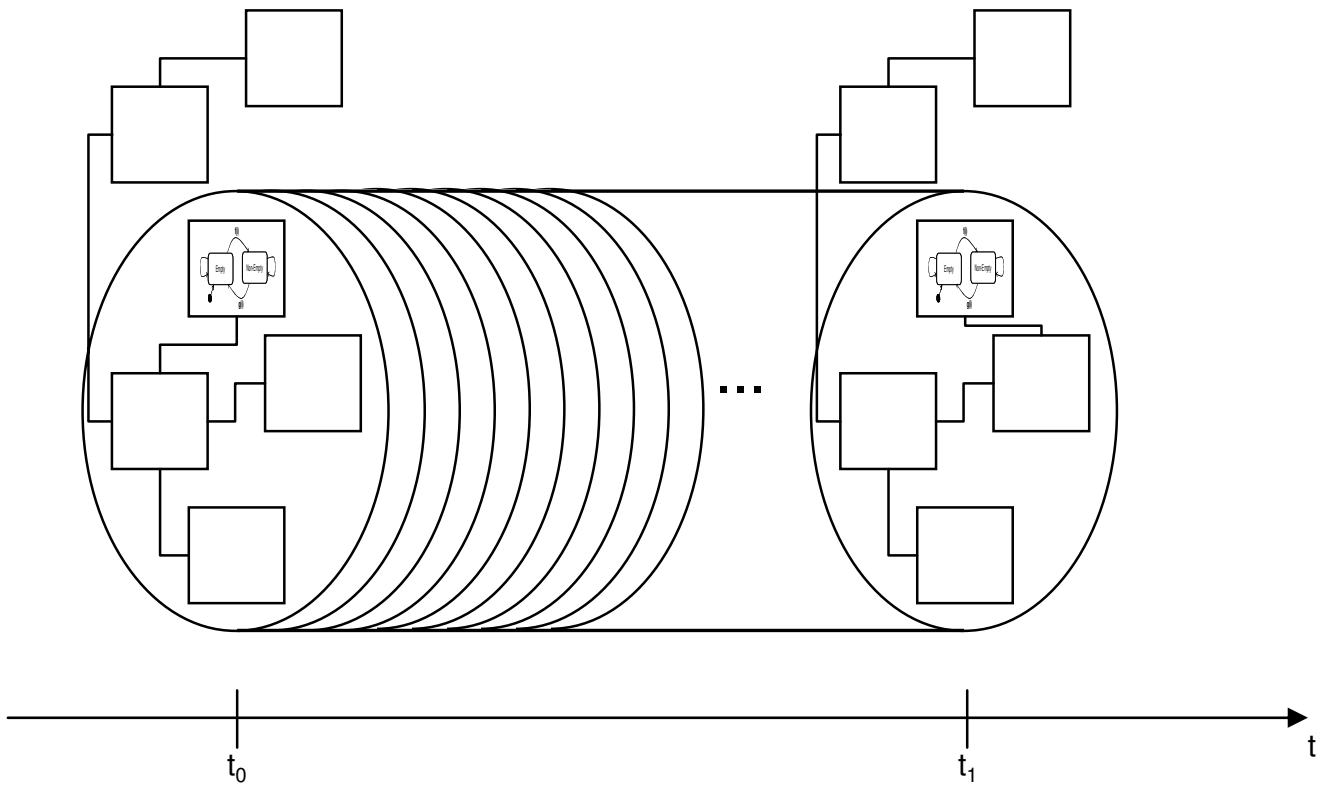

Figure 7.1: Multiple transitions of a TSTS for one Statechart transition

A system model $sys$ is a correct implementation of a Statechart $sc \in \mathit{SCSimp}$ if, for every object $o \in \mathsf{UOID}_{sys}$ with $\mathsf{classOf}(o) = \mathit{sc.className}$, there is a group of objects $oids \subseteq \mathsf{UOID}_{sys}$ with $o \in \mathit{oids}$ and a map

$$\Pi : \mathit{sc.states} \rightarrow \wp(\mathit{states}(\mathit{oids}))$$

that satisfies the following conditions:

1. The Statechart invariant holds for all states of the object group that are in the range of $\Pi$, regardless of a specific thread or valuation (indicated as _).

$$\forall \mathit{ogs} \in (\bigcup_{\forall \mathit{scs} \in \mathit{sc.states}} \Pi(\mathit{scs})) : \mathcal{C}(\mathit{sc.inv}, \mathit{ogs}, o, _, _)$$

2. States of the object group that correspond to initial states of the Statechart are initial states in the system model.

$$\forall \mathit{scs} \in \mathit{sc.states} : \mathit{scs.modifier} = \mathsf{initial} \implies \Pi(\mathit{scs}) \subseteq \mathit{init(oids)}$$

3. The state invariant holds for all states of the group of objects $ogs$ that correspond to a Statechart state $scs$.

$$\forall \mathit{scs} \in \mathit{sc.states}, \forall \mathit{ogs} \in \Pi(\mathit{scs}) : \mathcal{C}(\mathit{scs.inv}, \mathit{ogs}, o, _, _)$$

4. The system shows run-to-completion behavior. It does not contain states of the group of objects in which more than one state-relevant message is processed.

$$\begin{aligned}&\forall \mathit{ogs} \in \mathit{states}(\mathit{oids}) : \\ &(\nexists m_i \in \mathsf{callsOf}(o), m_1 \neq m_2, t_i \in \mathsf{UTHREAD}_{sys}\ (i = 1, 2) : \\ &\quad \mathsf{nameOf}(m_i) \in \mathit{triggers}(\mathit{sc}) \wedge \mathsf{proc}(\mathit{ogs}(o)(t_i), m_i))\end{aligned}$$

5. Let $t = (\mathit{scs}, \varphi_1, c, (\mathit{stmt}, \varphi_2), \mathit{sct}) \in \mathit{sc.transitions}$ be a Statechart transition, $\mathit{ogs} \in \Pi(\mathit{scs})$, and $th \in \mathsf{UTHREAD}$.

   The Statechart transition $t$ can be taken if it is enabled. To be enabled there has to be a matching message $m$ in the object's buffer for which a valuation $v$ can be found that binds all variables in the call $c$. Additionally, the pre-condition $\varphi_1$ must hold:

$$\begin{aligned}&\mathit{enabled} \iff \\ &\quad \exists m \in \mathsf{msg}(c, \mathit{buffer}(\mathit{ogs}, o), o) \wedge \\ &\quad \exists (v \in V) : \mathcal{E}(c, \mathit{ogs}, o, th, _, v) \wedge \\ &\quad \mathcal{C}(\varphi_1, \mathit{ogs}, o, th, v)\end{aligned}$$

   If the transition is enabled, it can be taken. One transition in the Statechart may correspond to many state changes in the object group state. That means that there exist $n$ state transitions of the object group that lead from the source state $ogs$ to the target state $os_3 \in \Pi(\mathit{sct})$. Some time in between there exist two object group states $og_1$ and $og_2$ that reflect the execution of the statement *stmt*. Finally the message $m$ has been consumed and the post-condition $\varphi_2$ must hold.

$$\begin{aligned}&\mathit{enabled} \implies \\ &\quad \exists os_1, os_2, os_3 \in \mathit{states}(\mathit{oids}), n \in \mathbb{N} : \\ &\quad os_i \in \pi_1(\Delta(o)^{n_i}(\mathit{ogs}), _), (i = 1, 2, 3), n_1 \leq n_2 \leq n_3 \\ &\quad os_3 \in \Pi(\mathit{sct}) \wedge \\ &\quad \mathcal{S}(\mathit{stmt}, os_1, o, t, v, os_2, _) \wedge \\ &\quad m \notin \mathsf{msg}(c, \mathit{buffer}(\mathit{ogs}, o), o) \wedge \\ &\quad \mathcal{C}(\varphi_2, os_3, o, th, v)\end{aligned}$$

Note that we do not have to resolve conflicting enabled transitions. For each enabled Statechart transition, transitions in the system model exist. The transition system in the system model then non-deterministically selects a transition To reduce non-determinism in the system model, additional constraints can be imposed.

To summarize, each system model that satisfies the above conditions is a valid realization of the Statechart specification. In general, there will be many system models that are possible realizations.

It should now become clear that we are also able to give semantics to UML specifications that, e. g., contain multiple Statecharts describing the same class. Thanks to the set-valued semantics, the overall semantics is established by the intersection of all system models that are the result of mapping single diagrams.

# 8 Mapping existing Statecharts Semantics into the System Model

## 8.1 Introduction

In this section we investigate how an existing semantics of statemachines can be mirrored in our framework. The idea is to reformulate a given semantics into a mapping of statemachines into system models. An arbitrary semantics of statemachines is most likely some kind of transition system. The existing semantics is then reflected by mapping the (abstract) states of its semantic domain to (concrete) states of a system model. This mapping is such that the images of abstract states reachable from a given one by means of the (abstract) transition function are reachable from the image of the given abstract state by means of the (concrete) transition function of the system model. More specifically, if $s_1$ and $s_2$ are abstract states and $s_2$ is reachable from $s_1$, then the image of $s_2$ in the system model is likewise reachable from the image of $s_1$. We demonstrate the procedure by making use of the semantics in [13] which uses Kripke structures.

## 8.2 Semantics of UML Statemachines

The mapping described in the present work is based on the semantics for statemachines of [13]. In the following a brief account of this approach is given. The advantages of the chosen semantics is that it supports not only interlevel transitions, as also e.g. [1, 2, 3] do, but also the history mechanism as well as entry/exit actions.

[13] starts by introducing an abstract syntax of UML statemachine terms. Let $\mathcal{N}$ be a set of *state names* and $\mathcal{T}$ be a set of *transition names*. Let $\mathbb{E}$ be a set of events and $\mathbb{A}$ be a set of actions. Let HT denote the set $\{\mathsf{none}, \mathsf{deep}, \mathsf{shallow}\}$. Let $\mathbb{N}_k$ denote the set $\{1, \ldots, k\}$ and $\mathsf{TR}(k)$ denote $\mathcal{T} \times \mathbb{N}_k \times \wp(\mathcal{N}) \times \mathbb{E} \times \mathbb{A}^* \times \wp(\mathcal{N}) \times \mathbb{N}_k \times \mathsf{HT}$, where $k \in \mathbb{N}$, $k > 0$.

1. If $n \in \mathcal{N}$ and $en, ex \in \mathbb{A}^*$, then $[n, (en, ex)]$ is a UML statemachine *basic* term.

2. If $n \in \mathcal{N}$, $s_1, \ldots, s_k$ are UML statemachine terms ($k > 0$), and $en, ex \in \mathbb{A}^*$, then $[n, (s_1, \ldots, s_k), (en, ex)]$ is a UML statemachine *and* term.

3. If $n \in \mathcal{N}$, $s_1, \ldots, s_k$ are UML statemachine terms ($k > 0$), $l \in \mathbb{N}_k$, $T \subseteq \mathsf{TR}(k)$, and $en, ex \in \mathbb{A}^*$, then $[n, (s_1, \ldots, s_k), l, T, (en, ex)]$ is a UML statemachine *or* term.

The set of all UML statemachine terms is denoted by UML-SM.

For a basic term $[n, (en, ex)]$, $n$ is its name, $en$ is its entry action, and $ex$ is its exit action. For an and-term $[n, (s_1, \ldots, s_k), (en, ex)]$, $s_1, \ldots, s_k$ are moreover its subterms. Further for an or-term $[n, (s_1, \ldots, s_k), l, T, (en, ex)]$, $s_l$ is its *active* subterm and $T$ the set of transitions relating its subterms.

Let $[n, (s_1, \ldots, s_k), l, T, (en, ex)]$ be an or-term. Let $(t, i, N_s, e, \alpha, N_t, j, \mathsf{ht}) \in T$ be one of the term's transitions. We say that $t$ is the name of the transition, $s_i$ is its source and $s_j$ its target, $e$ is its trigger and $\alpha$ its action, $N_s$ is its source restriction and $N_t$ its target determinator, and $\mathsf{ht}$ is its history type. Source restriction and target determinator provide means for expressing interlevel transitions; if either is non-empty, the transition connects different levels. Some well-formedness restrictions apply, namely that the involved state and transition names are all different as well as non-trivial ones concerning source restriction and target determinator; for details, the reader is referred to [13].

The cited work continues by defining a two level semantics. The first (auxiliary) level derives judgments of the form $s \xrightarrow[\alpha]{e}_{\mathrm{f}} s'$ where $s, s' \in \mathsf{UML\text{-}SM}$ are UML statemachine terms, $e \in \mathbb{E}$ is an event, $\alpha \in \mathbb{A}^*$ is a sequence of actions, and $\mathrm{f} \in \{0, 1\}$ is a flag.[1] The axiom schematas and inference rules for deriving these judgments are given in Tab. 8.1. Some auxiliary functions are needed. Intuitively, $\mathsf{conf}$ computes the current configuration, i.e., the set of names of all currently active substates of a given UML statemachine term. *entry* and *exit* return the set of all possible sequences of entry and exit actions, respectively, of a given UML statemachine term; these two functions respect the non-deterministic nature of the search they perform, which is mirrored by the fact that each of them returns not a single sequence but a set of sequences of actions. $\mathsf{next}$ computes the state which becomes active after a transition has been fired. The precise definition of these functions can be found in [13].

The relation $s \xrightarrow[\alpha]{e}_{\mathrm{f}} s'$, roughly speaking, states that the UML statemachine term $s$, given stimulus (or input) $e$, performs the actions in (or outputs) $\alpha$ and evolves to the UML statemachine term $s'$ with stuttering flag f. It might be confusing that in rule ($\mathrm{basic_{aux}}$), for instance, the entry and exit actions of the state are not output after the (stuttering) transition has been fired. The clue resides precisely in the fact that it is a stuttering step, i.e., just an event is consumed and the state does not change. Notice that a basic UML statemachine term contains no transitions; thus, no stimulus can fire the term's entry and exit actions. These are performed in a context where the basic UML statemachine term is reached or left by a transition in a UML statemachine term including the basic one. For a throughout discussion, the read should consult [13].

Now this relation is used to define a Kripke structure that explains the behavior of a UML statemachine term. Kripke structures are non-deterministic finite state machine usually used to represent the behavior of a system. They are basically a graph whose nodes represent the (reachable) states of the system and whose edges represent state transitions. They moreover have a distinguished subset of states, the initial ones. In our case, the nodes of the Kripke structure are pairs, a UML statemachine term and a sequence of events. The sequence of events represents the (external) input. The edges are not explicitly given but inductively defined in Fig. 8.1. More precisely, the inference system defined in Fig. 8.1 consists of only one axiom schema which allows the

[1] A positive flag $\mathrm{f} = 1$ indicates that the other elements of the judgment are in the derived relation because a transition of the UML statemachine term on the left is fired. A negative flag $\mathrm{f} = 0$ means that an event is consumed but no transition is fired, i.e., source and target UML statemachine terms are identical and the sequence of actions is empty. In this latter case we speak of a stutter step. Flags can be "or-ed", the result being 0 if all of them are negative, 1 otherwise.

$$(\text{basic}_{\text{aux}})\ \ [n,(en,ex)] \xrightarrow[\langle\rangle]{e}_0 [n,(en,ex)]$$

$$(\text{or}^1_{\text{aux}})\ \ [n,(s_1,\ldots,s_k),i,T,(en,ex)] \xrightarrow[e_1::\alpha::e_2]{e}_1$$

$$[n,(s_1,\ldots,s_k)[s_j \mapsto \mathsf{next}(\mathsf{ht},N_t,s_j)],j,T,(en,ex)]$$

if $(_,i,N_s,e,\alpha,N_t,j,\mathsf{ht}) \in T,\ N_s \subseteq \mathsf{conf}(s_i),$ and $s_i \not\xrightarrow{e}_1$

where $e_1 \in \textit{exit}(s_i)$ and $e_2 \in \textit{entry}(\mathsf{next}(\mathsf{ht},N_t,s_j))$

$$(\text{or}^2_{\text{aux}})\ \ \frac{s_l \xrightarrow[\alpha]{e}_1 s'_l}{[n,(s_1,\ldots,s_k),l,T,(en,ex)] \xrightarrow[\alpha]{e}_1 [n,(s_1,\ldots,s_k)[s_l \mapsto s'_l],l,T,(en,ex)]}$$

$$(\text{or}^3_{\text{aux}})\ \ \frac{s_l \xrightarrow[\langle\rangle]{e}_0 s_l}{[n,(s_1,\ldots,s_k),l,T,(en,ex)] \xrightarrow[\langle\rangle]{e}_0 [n,(s_1,\ldots,s_k),l,T,(en,ex)]}$$

if $[n,(s_1,\ldots,s_k),l,T,(en,ex)] \not\xrightarrow{e}_1$

$$(\text{and}_{\text{aux}})\ \ \frac{(s_j \xrightarrow[\alpha_j]{e}_{\mathrm{f}_j} s'_j)_{1\le j\le k}}{[n,(s_1,\ldots,s_k),(en,ex)] \xrightarrow[\alpha]{e}_{\mathrm{f}} [n,(s'_1,\ldots,s'_k),(en,ex)]}$$

where $\mathrm{f} = \bigvee_{j=1}^{k} \mathrm{f}_j$ and $\alpha \in \{\alpha_{\pi(1)} :: \cdots :: \alpha_{\pi(k)} \mid \pi$ is a permutation of size $k\}$

Table 8.1: Auxiliary semantics

(consume-input) $(s_1, \epsilon_1) \Rrightarrow (s_2, \epsilon_2)$

if $s_1 \xrightarrow[\alpha]{e}_{\mathsf{f}} s_2$

and there exists $\epsilon'$ such that $(\epsilon_1, e, \epsilon') \in \mathsf{sel}$ and $(\alpha, \epsilon', \epsilon_2) \in \mathsf{join}$ derivation of judgments of the form $(s_1, \epsilon_1) \Rrightarrow (s_2, \epsilon_2)$. If one such judgment can be inferred, then there is an edge from $(s_1, \epsilon_1)$ to $(s_2, \epsilon_2)$ in the Kripke structure.

Figure 8.1: Operational semantics of UML statemachines

Here, again, some auxiliary relations are needed. These are sel and join. Whereas sel determines if a sequence results to another sequence after removing one of its elements, join does somehow the opposite as it determines if a sequence is the result of adding an element to another sequence. These relations are deliberately left unspecified, since they establish a scheduling strategy of the input event queue.

Now it becomes apparent why in [13] necessarily the set of events and the set of actions must be the same. That is, $\mathbb{E} = \mathbb{A}$, since the actions output by a step form part of input of the next step.

## 8.3 Semantic Mapping

In the present section we study the relationship between a UML statemachine and a system model. While system models describe concrete system implementations, UML statemachines on the contrary provide a means to abstractly depict what a (part of a) system which traverses different states does in each state and/or how it reacts to stimuli steming from its context. Thus a natural choice for relating these two views of the same system is to search for a map from abstract to concrete states, i.e., from the states of a UML statemachine to the states of a system model.

A closer look at the nature of both languages induces the restriction of the image of that map to the states of a particular object. Indeed, a UML statemachine typically describes the behavior of an object of a particular class, whereas an implementation as a whole is the main concern of a system model. Therefore, we consider the abstract states and their correspondent concrete ones. That is, given a UML statemachine term for objects of a class, and an instance $o$ of that class in a system model, the projection function maps an abstract state (of the UML statemachine term) to a set of concrete states (of the instance $o$):

$$\Pi : \mathsf{UML\text{-}SM} \times o : \mathsf{INSTANCE} \to \wp(\mathit{states}(o))$$

where INSTANCE is the set of all instances, and $\mathsf{classOf}(o)$ returns the class name of interest; see [1].

The projection function $\Pi$ of above identifies a number of concrete states of an instance of a particular class that correspond to an abstract one of any object of that class. These correspon-dence has a counterpart regarding transitions:

if $(s_1, \epsilon_1) \Rrightarrow (s_2, \epsilon_2)$, then $\forall \mathsf{st} \in \Pi(s_1).\ \exists n \in \mathbb{N}.\ \exists \mathsf{st}_1, \ldots, \mathsf{st}_n \in \mathit{states}(o)$ :
$\mathsf{st}_1 \in \pi_1(\Delta.o(\mathsf{st}, _)) \wedge$
$\mathsf{st}_2 \in \pi_1(\Delta.o(\mathsf{st}_1, _)) \wedge \ldots \wedge$
$\mathsf{st}_n \in \pi_1(\Delta.o(\mathsf{st}_{n-1}, _)) \wedge$
$\mathsf{st}_n \in \Pi(s_2)$

This means, a macrostep on the specification level corresponds to a series of microsteps on the implementation level.

It remains to investigate the relationship between the input queues $\epsilon_1$, $\epsilon_2$ and the successive event stores of the series of microsteps. This is done in the next section using an example.

## 8.4 Example

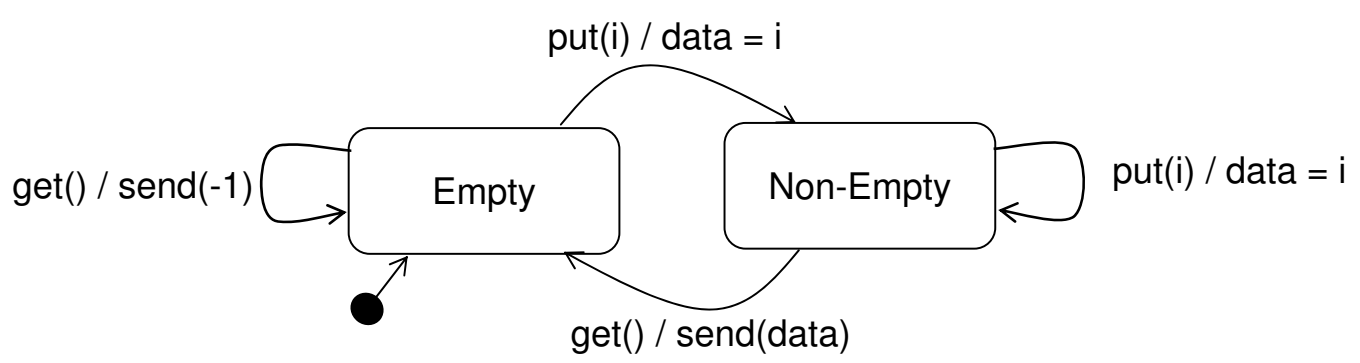

Figure 8.2: Example statemachine of a Buffer

The example statemachine in Fig. 8.2 is expressed in terms of statemachine terms as follows:

$$\begin{aligned}
empty &= [\mathsf{Empty}, (\langle\rangle, \langle\rangle)] \\
nonEmpty &= [\mathsf{NonEmpty(v)}, (\langle\rangle, \langle\rangle)] \\
buffer &= [\mathsf{Buffer}, (empty, nonEmpty), 1, t, (\langle\rangle, \langle\rangle)] \\
t &= \{(t_1, 1, \emptyset, get(), \langle send(-1)\rangle, \emptyset, 1, \mathsf{none}), \\
&\quad (t_2, 1, \emptyset, put(i), \langle\rangle, \emptyset, 2(i), \mathsf{none}), \\
&\quad (t_3, 2(v), \emptyset, get(), \langle send(v)\rangle, \emptyset, 1, \mathsf{none}), \\
&\quad (t_4, 2(v), \emptyset, put(i), \langle\rangle, \emptyset, 2(i), \mathsf{none}) \qquad \}
\end{aligned}$$

Since statemachine terms do not explicitly support variables that are needed to store the value of the data received by the buffer, we encode that in the name of the state *nonEmpty*. $v$ may be seen as the current value that the buffer holds.

A concrete run of the state machine for the input sequence $\langle put(3), get()\rangle$ is given by a run of the corresponding Kripke structure:

$$\begin{aligned}
&([\mathsf{Buffer}, (empty, nonEmpty(_)), 1, t, (\langle\rangle, \langle\rangle)], \langle put(3), get()\rangle) \Rrightarrow \\
&([\mathsf{Buffer}, (empty, nonEmpty(3)), 2, t, (\langle\rangle, \langle\rangle)], \langle get()\rangle) \Rrightarrow \\
&([\mathsf{Buffer}, (empty, nonEmpty(_)), 1, t, (\langle\rangle, \langle\rangle)], \langle send(3)\rangle)
\end{aligned}$$

where sel and join are assumed to behave in a FIFO-fashion.

System models describe possible system realizations. There are many ways to define a system that implements the behavior of the statemachine. Thus, the projection $\Pi$ is not unique. In the present example and given one such system, we assume that the states of an object are determined by the value of *data*, i.e., we assume a partition of the state space of the object depending on the value of *data*. Other realizations may follow other strategies, for instance a special attribute of the object may be used to determine the current state of the object.

The projection $\Pi$ from the UML statemachine terms to the states of an object in a system model can thus be defined as

$$\begin{aligned}\Pi([\mathsf{Empty}, (\langle\rangle, \langle\rangle)]) &= \{state(s,o) | s \in \mathsf{USTATE} \wedge val(ds(s), o, data) = -1\} \\ \Pi([\mathsf{NonEmpty(v)}, (\langle\rangle, \langle\rangle)]) &= \{state(s,o) | s \in \mathsf{USTATE} \wedge val(ds(s), o, data) = v\} \\ \Pi(\mathit{buffer}) &= \Pi(\pi_{\pi_3(\mathit{buffer})}(\pi_2(\mathit{buffer})))\end{aligned}$$

The last definition is the general one for or-terms: given an or-term $[n, (s_1, \ldots, s_k), l, T, (en, ex)]$, its *active* subterm is $s_l$ and thus the projection $\Pi$ on the or-term is the projection of the $l$-th subterm in the list of subterms. While the number $l$ is the third component of the or-term (i.e., its $\pi_3$ component), the list of subterms is its second component (i.e., its $\pi_2$ component). That is, the projection $\Pi$ of the or-term is the projection $\Pi$ of the $\pi_3$-th component (i.e., the $\pi_{\pi_3(_)}$ component) of the second component: $\Pi(s) = \Pi(\pi_{\pi_3(s)}(\pi_2(s)))$ if $s$ is an or-term.

Fig. 8.3 illustrates the relation of the Kripke structure (macrostep level) and the series of microsteps of a system model. The system model state $s_1$ is identified with the state *NonEmpty*. The same applies to the system model state $s_6$ and the state *Empty*. In between, a series of microsteps is executed that reveal the implementation decisions of using a local variable $t$ to store the current value after which the *data* value is set to $-1$ in order to establish a correct state partition.

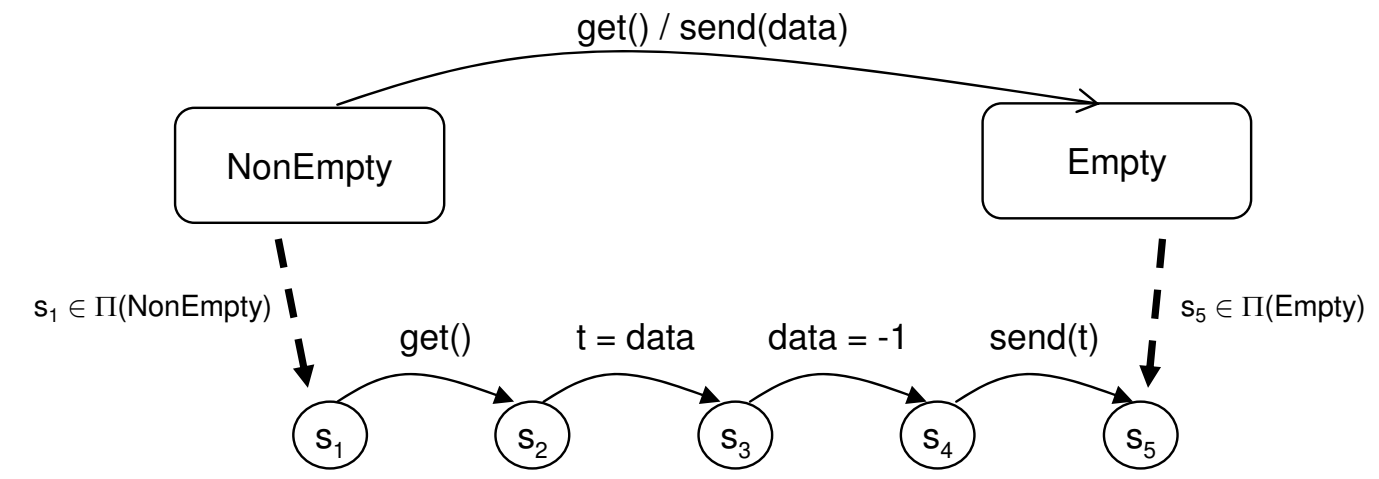

Figure 8.3: Microsteps of a system model that fulfills the statemachine

Formally, there exist $\mathsf{st}_1 \in \Pi([\mathsf{NonEmpty(v)}, (\langle\rangle, \langle\rangle)])$ and $\mathsf{st}_2 \in \pi_1(\Delta.o(\mathsf{st}_1, \langle m_{get}\rangle))$ where

$$\begin{aligned} m_{get} &\in \mathit{signalsOf}(o, sg, o') \subseteq \mathit{msgIn}(o) \subseteq \mathsf{UMESSAGE} \\ \mathit{nameOf}(sg) &= mathitget \\ \mathit{classOf}(sg) &= \mathit{classOf}(o) \\ \text{and } \mathit{parTypes}(sg) &= \langle\rangle \end{aligned}$$

for a sender object $o'$, such that further conditions hold. These conditions depend on the nature of the communication between the object of interest, a buffer, and its clients. We assume that the communication is asynchronous, and introduce a function $\mathit{signalsOf}(oid_1, sg, oid_2)$ in the style of $callsOf$; cf. [2, p. 15]. Let us go into the details of the additional conditions for the projection function $\Pi$.

Let $(\mathsf{st}_i, M_i) \in \Delta.o(\mathsf{st}_{i-1}, _)$ for $i > 2$ be a run of the system, and let $k$ be the smallest index with $\mathsf{st}_k \in \Pi([\mathsf{Empty}, (\langle\rangle, \langle\rangle)])$. Then, there exists exactly one index $s$ with $2 \leq s \leq k$ with $M_s = \langle m_{send} \rangle$ such that

$$\begin{aligned} m_{send} &\in signalsOf(o', sg', o) \subseteq msgOut(o) \subseteq \mathsf{UMESSAGE} \\ nameOf(sg') &= get \\ classOf(sg') &= classOf(o') \\ \text{and } parValues(m_{send}) &= \langle v \rangle \end{aligned}$$

where $parValues(m) = \pi_3(m)$ for any message $m$; cf. [2, p. 15].

If such index $k$ does not exist, then the (concrete) run is not a valid implementation of the above (abstract) run in the Kripke structure. Likewise, if the index $k$ exists, and an index $s$ does not exist or it is not unique, then the (concrete) run is not a valid implementation of the above (abstract) run in the Kripke structure.

A sequence of microsteps that is not a valid sequence for the abstract transition from *NonEmpty* to *Empty* is shown in Fig. 8.4. It is invalid because sending a message twice (with potentially different values) is not covered by the abstract transition in the Kripke structure that states that exactly one event is produced as the result of taking the transition.

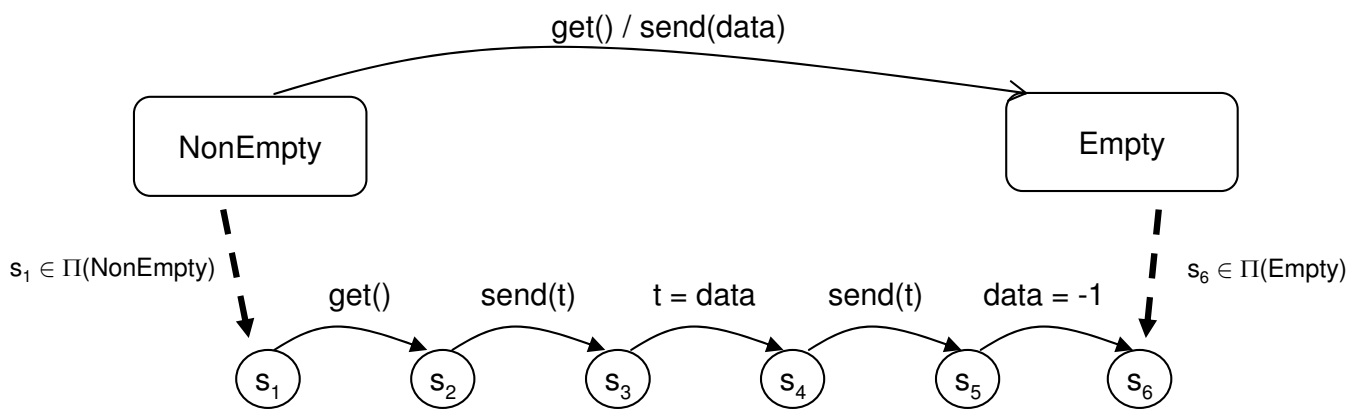

Figure 8.4: Microsteps of a system model that violates the statemachine

## 8.5 Semantic Mapping (contd.)

The requirement $\mathbb{E} = \mathbb{A}$ of [13] needs not hold in our setting. This is because a system model for a complete specification can be obtained by composition of system models for fragments of that specification. In this case, the set of events that can be input to a statemachine and the set of actions that the statemachine may output need not coincide. Notice, however, that this only affects the sel and join functions. These functions were left unspecified, what perfectly suits our purposes. sel selects the stimulus to be handled next, according to a desired scheduling strategy; it needs not select an element in $\mathbb{A} \setminus \mathbb{E}$. join adds an element to the input event queue; likewise, it can simply ignore an element in $\mathbb{A} \setminus \mathbb{E}$. This possibility, however, should be avoided since an output event usually is not to be simply discarded.

In favor of maintaining the requirement $\mathbb{E} = \mathbb{A}$ one can argue that those universes, $\mathbb{E}$ and $\mathbb{A}$, can be regarded as containing all possible stimuli, including those that are not pertinent for the specification fragment being considered. Doing so, however, would imply that this fragment performs stutter steps for input that is to be handled by other fragments. This not only means overhead, given that all statemachines (but possibly one) are performing a stutter step for each possible input. It might also mean that the very same input stimulus is consumed by two statemachines, if one message meant for an instance is not prevented from being delivered (also) to other instances.

In general, thus, a careful treatment of events is imperative. Events contained in the event store must be appropriately delivered to and consumed (or handled) by the corresponding class instance. Hence, it seems more suitable that statemachines take care of exactly those events they are supposed to react to. By the same token, the actions they emit are also to be appropriately delivered. In this respect, the system model definition is quite liberal, and when necessary should be strengthened.

Let $o$ be an object whose behavior is specified by $s_1$. Suppose $(s_1, \epsilon_1) \Rrightarrow (s_2, \epsilon_2)$, where $s_1 \xrightarrow[\alpha]{e}_{\mathrm{f}} s_2$ with $(\epsilon_1, e, \epsilon') \in \mathsf{sel}$ and $(\alpha, \epsilon', \epsilon_2) \in \mathsf{join}$ for some $\epsilon'$. Assume moreover there is $\mathsf{st} \in \Pi(s_1) \subseteq \mathit{states}(o) \subseteq \mathsf{USTATE}$. The state $\mathsf{st}$ is therefore a triple $(vals(ds, o), cs(o), es(o))$, where $(ds, cs, es)$ is a state in the universe $\mathsf{USTATE}$ of states of the system, $ds$ is a data store, $cs$ is a control store, and $es$ is an event store, from which attribute values, thread information, and events, respectively, associated with an object can be retrieved. Let $(\mathsf{st}_i, M_i)$ $(i \geq 0)$ be a run for $o$ of the system from state $\mathsf{st}$, i.e., $\mathsf{st}_0 = \mathsf{st}$ and $(\mathsf{st}_j, M_j) \in \Delta.o(\mathsf{st}_{j-1}, _)$ $(j > 0)$. If there exist indices $k$, $r$ and $s$ $(0 \leq r \leq s \leq k)$ such that

1. $\mathsf{st}_k \in \Pi(s_2)$
2. $\mathsf{st}_r$ contains $e$ as input event for $o$
3. $\mathsf{st}_{r+1}$ does not contain $e$ as input event for $o$
4. $\alpha \in M_S$ if $\alpha$ is an output event of $o$
5. there are no indices between $0$ and $k$, different from $r$ $(s)$, fulfilling conditions 2 and 3 (condition 4),

then the run satisfies the transition of the statemachine term from $s_1$ to $s_2$.

The approach presented above is viable for further semantics of statemachines. The procedure followed can also be suitably applied to other UML sublanguages. Indeed, the idea of defining a mapping from an abstract language to a (number of) concrete states, in such a way that some semantic restriction holds, seems promising. Possible candidates are the UML sublanguages of interactions and of activities, since they may be provided with trace-based semantics.

# 9 Related Work and Conclusion

In this report, we have defined the semantics of UML/P Statecharts using the system model as the semantic domain. Additionally, we investigated how an arbitrary Statechart semantics can be reformulated in our system model allowing us to reuse other work when defining an integrated semantics of UML.

We briefly state related work. An extensive review of related work on Statechart semantics can be found in [5]. Compared to most other approaches, the main difference of our semantics is that it does not focus on firing rules and evolving Statechart configurations, but directly associates possible realizations to a Statechart behavior in terms of object oriented systems in the system model. [6] also defines the semantics of State Machines by transformation to a simplified version (core state machines) and then provides an operational semantics that characterizes the configuration steps in a State Machine run. The computational model used in [12] are hierarchical state machines (HTS) and a template-semantics is given to express the execution semantics (operational semantics) of State Machines while explicating the semantics choices that have been left open in the UML standard. Compared to these and other operational semantics definitions our denotational semantics has the advantage that it characterizes the behavioral properties that objects or groups of objects must have but does not exclude other behavior, thus allowing underspecification.

### Evaluation

The approach of defining the semantics for UML described in [4] was employed and found suitable for defining a semantics of UML/P Statecharts. It involved a transformation on the syntactic domain to reduce the number of language constructs and a semantic mapping for simplified Statecharts. The concise transformation rules given in this report are in the process of being implemented as transformations in the MontiCore framework to test and validate the individual transformation rules.

By reducing the number of syntactic constructs significantly (the abstract syntax of Full UML/P Statecharts has about 13 productions while the simplified abstract syntax of Statecharts has 7 productions), we are able to define the semantics of these constructs purely on the syntactic domain which should be a bit easier accessible than the mapping to the system model. The mapping itself also benefits from the syntax transformations because fewer constructs have to be mapped, so in general the mapping should be more comprehensible than a mapping for Full Statecharts. Since the transformations were defined on the mathematical abstract syntax, a transformation scheme was developed that allows for a comprehensible specification of transformation rules.

Because of the modular language design that uses embedding of languages to state actions and conditions, the semantics mapping is further simplified, assuming the semantics for the embedded language is given.

The system model is defined using maths. Consequently, it was straightforward to define additional mathematical machinery that can be used in the semantics mapping. States of object groups as well as a notion of reachability for object group states were added as system model extensions.

A systematic treatment of semantic variation points is still missing. Semantic variation points are for example explicitly present in the syntax as stereotype (e.g., to determine priorities of conflicting transitions) or are left unspecified in the semantics mapping (choice of next input to process). This is a matter of future work.

Technische Universität Braunschweig
Informatik-Berichte ab Nr. 2003-10

| | | |
|---|---|---|
| 2003-10 | M. Mutz, M. Huhn | Automated Statechart Analysis for User-defined Design Rules |
| 2004-01 | T.-P. Fries, H. G. Matthies | A Review of Petrov-Galerkin Stabilization Approaches and an Extension to Meshfree Methods |
| 2004-02 | B. Mathiak, S. Eckstein | Automatische Lernverfahren zur Analyse von biomedizinischer Literatur |
| 2005-01 | T. Klein, B. Rumpe, B. Schätz (Herausgeber) | Tagungsband des Dagstuhl-Workshop MBEES 2005: Modellbasierte Entwicklung eingebetteter Systeme |
| 2005-02 | T.-P. Fries, H. G. Matthies | A Stabilized and Coupled Meshfree/Meshbased Method for the Incompressible Navier-Stokes Equations — Part I: Stabilization |
| 2005-03 | T.-P. Fries, H. G. Matthies | A Stabilized and Coupled Meshfree/Meshbased Method for the Incompressible Navier-Stokes Equations — Part II: Coupling |
| 2005-04 | H. Krahn, B. Rumpe | Evolution von Software-Architekturen |
| 2005-05 | O. Kayser-Herold, H. G. Matthies | Least-Squares FEM, Literature Review |
| 2005-06 | T. Mücke, U. Goltz | Single Run Coverage Criteria subsume EX-Weak Mutation Coverage |
| 2005-07 | T. Mücke, M. Huhn | Minimizing Test Execution Time During Test Generation |
| 2005-08 | B. Florentz, M. Huhn | A Metamodel for Architecture Evaluation |
| 2006-01 | T. Klein, B. Rumpe, B. Schätz (Herausgeber) | Tagungsband des Dagstuhl-Workshop MBEES 2006: Modellbasierte Entwicklung eingebetteter Systeme |
| 2006-02 | T. Mücke, B. Florentz, C. Diefer | Generating Interpreters from Elementary Syntax and Semantics Descriptions |
| 2006-03 | B. Gajanovic, B. Rumpe | Isabelle/HOL-Umsetzung strombasierter Definitionen zur Verifikation von verteilten, asynchron kommunizierenden Systemen |
| 2006-04 | H. Grönniger, H. Krahn, B. Rumpe, M. Schindler, S. Völkel | Handbuch zu MontiCore 1.0 - Ein Framework zur Erstellung und Verarbeitung domänenspezifischer Sprachen |
| 2007-01 | M. Conrad, H. Giese, B. Rumpe, B. Schätz (Hrsg.) | Tagungsband Dagstuhl-Workshop MBEES: Modellbasierte Entwicklung eingebetteter Systeme III |
| 2007-02 | J. Rang | Design of DIRK schemes for solving the Navier-Stokes-equations |
| 2007-03 | B. Bügling, M. Krosche | Coupling the CTL and MATLAB |
| 2007-04 | C. Knieke, M. Huhn | Executable Requirements Specification: An Extension for UML 2 Activity Diagrams |
| 2008-01 | T. Klein, B. Rumpe (Hrsg.) | Workshop Modellbasierte Entwicklung von eingebetteten Fahrzeugfunktionen, Tagungsband |
| 2008-02 | H. Giese, M. Huhn, U. Nickel, B. Schätz (Hrsg.) | Tagungsband des Dagstuhl-Workshop MBEES: Modellbasierte Entwicklung eingebetteter Systeme IV |
| 2008-03 | R. van Glabbeek, U. Goltz, J.-W. Schicke | Symmetric and Asymmetric Asynchronous Interaction |
| 2008-04 | M. V. Cengarle, H. Grönniger B. Rumpe | System Model Semantics of Statecharts |
| 2008-05 | M. V. Cengarle, H. Grönniger B. Rumpe | System Model Semantics of Class Diagrams |